# Supervised Denoising of Diffusion-Weighted Magnetic Resonance Images Using a Convolutional Neural Network and Transfer Learning

Jakub Jurek[*1], Andrzej Materka[1] || Kamil Ludwisiak[2], Agata Majos[3] || Kamil Gorczewski[4], Kamil Cepuch[4], Agata Zawadzka[4]

[1] Institute of Electronics, Lodz University of Technology, Aleja Politechniki 10, PL-93590 Lodz, Poland

[2] Department of Diagnostic Imaging, Independent Public Health Care, Central Clinical Hospital, Medical University of Lodz, Pomorska 251, PL-92213 Lodz, Poland

[3] Department of Radiological and Isotopic Diagnosis and Therapy, Medical University of Lodz, Pomorska 251, PL-92213 Lodz, Poland

[4] Siemens Healthcare GmbH, Henkestr. 127, 91052 Erlangen, Germany

**\*corresponding author**: jakub.jurek@p.lodz.pl

# Abstract

*In this paper, we propose a method for denoising diffusion-weighted images (DWI) of the brain using a convolutional neural network trained on realistic, synthetic MR data. We compare our results to averaging of repeated scans, a widespread method used in clinics to improve signal-to-noise ratio of MR images. To obtain training data for transfer learning, we model, in a data-driven fashion, the effects of echo-planar imaging (EPI): Nyquist ghosting and ramp sampling. We introduce these effects to the digital phantom of brain anatomy (BrainWeb). Instead of simulating pseudo-random noise with a defined probability distribution, we perform noise scans with a brain-DWI-designed protocol to obtain realistic noise maps. We combine them with the simulated, noise-free EPI images. We also measure the Point Spread Function in a DW image of an AJR-approved geometrical phantom and inter-scan movement in a brain scan of a healthy volunteer. Their influence on image denoising and averaging of repeated images is investigated at different signal-to-noise ratio levels. Denoising performance is evaluated quantitatively using the simulated EPI images and qualitatively in real EPI DWI of the brain. We show that the application of our method allows for a significant reduction in scan time by lowering the number of repeated scans. Visual comparisons made in the acquired brain images indicate that the denoised single-repetition images are less noisy than multi-repetition averaged images. We also analyse the convolutional neural network denoiser and point out the challenges accompanying this denoising method.*

# 1. Introduction

## 1.1. Theoretical background and study motivation

Diffusion-weighted magnetic resonance imaging (DW MRI, DWI) plays an important role in medicine and biology. It is one of the primary tools in studying brain function and visualizing abnormalities such as tumours or brain stroke [Lutsep1997, Messina2020]. Diffusion is a microscale non-bulk motion that occurs at short time scales, and therefore it requires rapid imaging sequences. DW images are thus most commonly acquired using Echo-Planar Imaging (EPI), a technique that allows to acquire whole image slices within times below a second [Mansfield1977]. However, the nature of the applied imaging sequences causes DW imagery to be more blurry and significantly more noisy than, for example, conventional spin-echo images [Johansen-Berg]. This is partly due to the fact that image contrast in DWI is obtained on the basis of attenuating nuclear magnetic resonance (NMR) signals originating from protons that diffuse, while the noise level in the receiver, which mixes with the received signals, does not change. EPI is also used in other MRI areas, such as the functional MRI of the brain, where the blood-oxygenation-level-dependent signal is used to create activation maps of the brain. Noise was reported to be an issue in functional brain MRI and forced time-consuming repeated acquisitions in a recent study [Himmelberg2021].

In DWI, a factor related to the diffusion-sensitizing magnetic field gradients, called the b-value, defines the amount of diffusion-weighting. Signal attenuation is exponentially related to the b-value. In a typical DWI experiment, images are collected with multiple b-values (e.g. 0-300-1000 for brain imaging). Image contrast observed in DW images is due to variations of multiple tissue properties like proton density, T2 and T1 relaxation times and the apparent diffusion coefficient (ADC). Locally, T2-related contrast may overwhelm diffusion-related contrast. To minimize the influence of other contrast-generating tissue properties, it is common to estimate ADC maps and interpret them along with DW images. ADC maps can be estimated from at least two DW images acquired with different b-values using linear regression. Noise in DW images introduces bias to estimation of ADC and other diffusion parameters [Tax2022].

Many MRI acquisition protocols, and these related to DWI especially, involve repeated acquisition of the same imaging scene with constant imaging parameter settings. The goal of this procedure is to improve the signal-to-noise ratio (SNR) by averaging the repeated images. The number of repetitions is commonly referred to as the number of signal averages or the number of excitations (NEX). Repeated acquisition and averaging is a basic means of decreasing the noise level while keeping the resolution unchanged, but has several drawbacks. Firstly, patients are likely to move during and between scans. Movement can be corrected in post-processing, but this requires resampling that introduces additional blurring and errors. Secondly, averaging assumes that image noise has zero mean value. This assumption is met for NMR signals received by scanner coils (raw k-space data) and for complex-domain images that are obtained via the Fourier Transform of the demodulated NMR signals. However, complex-domain DW images acquired with non-zero b-values suffer from random phase variations between each scan, meaning that the real and complex components change in every imaging repetition, even though their absolute value remains the same, under the assumption of no patient movement [Jones2010]. Averaging complex-domain images in DWI is thus associated with the risk of signal cancellation and instead, modulus images are first computed from complex ones and then averaged. Computation of modulus images changes the original zero-mean Gaussian noise to Rician noise. The latter distribution converges to Gaussian as SNR grows and to Rayleigh's when SNR drops. In low-SNR regions, the intensity of averaged modulus images will be overestimated, because of non-zero-mean Rician noise [Gudbjartsson1995].

### 1.2. Previous research on DWI denoising

Diffusion-weighted imaging remains one of the actively researched topics. A recent review summarizes the state of the art on DWI pre-processing, including denoising, with special interest in brain DWI [Tax2022]. In this review, the authors divided denoising methods into smoothing-, regularized-, wavelet-, PCA- and neural-network-based denoising. Several works are cited and described in more details therein. Here, we focus on analysis of two previous works, that go in the similar direction to ours.

The first one is the application of the neural network training strategy called Noise2Self [Batson2019] to DW images by Fadnavis et al. [Fadnavis2020]. Noise2Self allows to train a denoising neural network in an unsupervised fashion, without the need for noise-free target images, thanks to certain statistical principles. It was developed shortly after the discovery of another strategy called Noise2Noise [Lehtinen2018]. In the latter, both input and target images are noisy and are instances of the same image, differing just by random noise. The assumption in Noise2Noise is that noise is zero-mean, which is not met for modulus DWI. Also, since Noise2Noise requires repeated noisy zero-mean images, it cannot be directly applied to learn from high-b-value DWI, due to phase variations between repetitions.

Fadnavis et al. [Fadnavis2020] tested their Patch2Self method in real and simulated DWI. For the latter case, they used the mean squared error and the $R^2$ metrics and compared denoising results to the clean ground truth. For real DWI, they evaluated the Fiber Bundle Coherency in an tractographic experiment.

The second relevant work, by Muckley et al., attempted to use neural networks for a simultaneous reduction of Gibbs ringing and noise in DW images [Muckley2020]. The networks were trained on synthetic non-MRI images with simulated ringing artifacts and noise. Within evaluation, the group studied image content removed by denoising (often referred to as the residuals).

Although Noise2Noise and Noise2Self are highly interesting and perhaps revolutionary ideas in image denoising, their main benefit is that clean target images, not possible to obtain in many imaging domains including MRI, are not necessary for training. However, conventional supervised learning with clean target images still outperforms Noise2Noise and Noise2Self in denoising quality [Batson2019]. Transfer learning is an alternative solution both for conventional supervised learning and for strategies like Noise2Noise or Noise2Self. In the MRI context, transfer learning allows to perform conventional supervised learning using noisy and noise-free image pairs from a domain similar to MRI. In particular, and in contrast to Muckley et al. [Muckley2020], we choose simulated MRI, where clean images can be obtained for the purpose of neural network training and evaluation. We focus on a single effect only:

the noise. To keep the training images possibly close to the real DWI, we build a model of EPI data with the use of a digital brain phantom and introduce typical EPI imaging effects in a data-driven fashion. We then conduct several experiments on this synthetic MRI dataset and also on real DWI images of the head.

In many image processing and analysis applications, including medical image processing, convolutional neural networks (CNNs) showed superior performance to traditional, non-learning-based methods. Instead of comparing our results to previous achievements such as PCA-based denoising [Veraart2016], we focus on the comparison to the method that is currently dominating in clinical imaging – averaging of repeatedly acquired images. Although widespread, it is limited in several aspects mentioned above and most importantly, it is time-consuming. We evaluate the results using quantitative metrics and visually compare single-repetition denoised DW images to ones obtained with larger NEX values. Moreover, we perform an analysis of the noise maps estimated by the CNN to learn its advantages and drawbacks.

## 2. Materials

### 2.1. BrainWeb dataset

The BrainWeb phantom is a digital map of brain anatomy with 181x217x181 voxel count and 1 mm x 1 mm x 1 mm spatial resolution, available publicly [BrainwebPage, Cocosco1997, Collins1998, Kwan1996, Kwan1999]. The map describes the spatial distribution of multiple tissue types and the proportional contribution of each tissue to the volume of each voxel. T1, T2 and proton density values are provided with the phantom.

### 2.2. DWI dataset

Our imaging experiment was appropriately approved at the Medical University of Lodz. Nine volunteers gave written consent to participate and underwent a DW imaging exam. We also collected phantom (AJR-approved) images and air scans (performed with empty scanner bore). DW images were collected using the Siemens Avanto 1.5 Tesla MRI machine with 2D Single-Shot Echo-Planar Imaging (Siemens *ep2diff*) at the Central Clinical Hospital, Lodz Medical University, Poland. We modified the standard local head DWI protocol to take into account the goals of our study. In particular, we used TR=5700 ms, TE = 114 ms, 5 mm slices thickness, 1.4 mm x 1.4 mm in-plane voxel size, FOV = 230 mm x 230 mm, matrix = 160 x 160, bandwidth = 1250 Hz/px, echo spacing = 0.89 ms, EPI factor = 160, Partial Fourier in the phase-encoding direction (6/8 lines), three orthogonal diffusion directions, three b-values (0, 300, 1000). Scan time resulted to be 21 min 24 sec with these parameters. As a rule, we turned off parallel imaging (Siemens iPAT) and all optional filters, but the system required the use of Partial Fourier imaging. NEX was set to the maximum allowed value, 32, for patient and phantom scanning and to 4 for air scanning.

Images were reconstructed from k-space using the method described in Section 3.1.

## 3. Methods

### 3.1. Echo-planar imaging and EPI image reconstruction

MR data are collected in the frequency domain, called the k-space. The relation between the measured NMR signal $S(t)$, equivalent to the k-space $K(k_x,k_y)$ and the complex image slice $I_c$ is given by [Johansen-Berg, Chap. 2, p. 14]:

$$S(t) = K(k_x, k_y) = \iint I_c(x,y) e^{j2\pi[k_x x + k_y y]} dx dy \qquad \text{Equation 1}$$

$$k_x = \gamma \int G_x(t) dt; \; k_y = \gamma \int G_y(t) dt \qquad \text{Equation 2}$$

In Equation 1 and 2, $k_x$ and $k_y$ are coordinates in the k-space, the domain of the demodulated NMR signals and x and y are coordinates in the image domain. Individual values of $k_x$ and $k_y$ depend on the waveform (often trapezoidal) of gradient magnetic fields $G_x$ and $G_y$ that are used to encode spatial location of the components of the received signals, and the gyromagnetic ratio $\gamma$, a nucleus-specific physical constant. The time coordinate t is mapped into the k-space via Equation 2, thus S can be organized in a 2D k-space matrix K. Rows of the k-space, lying in the X (frequency-encoding) direction, are often called k-space lines and they contain signals acquired during single phase-encoding steps. Direction Y is called the phase-encoding direction. The dimensions of the k-space match the dimensions of the complex image slice $I_c$. Due to various effects, the complex intensity values of the image have different phase, which is agreed to be smoothly varying [Pizzolato2020]. The EPI acquisition process is repeated for each slice.

Image slice $I_c$, as it can be derived from Equation 1, can be reconstructed by the inverse Fourier transform of the k-space. In EPI, however, this requires additional corrections due to effects described below. EPI is a rapid imaging technique, in which the whole k-space data matrix is acquired at once, within a single excitation pulse. Sampling of the k-space is equivalent to sampling of the NMR signal S(t) which is emitted by the electromagnetically excited protons as they relax to the equilibrium state. To decrease acquisition time, ramp sampling is performed, i.e. the samples are acquired while $G_x$ rises, saturates and then drops. This, in accordance with Equation 2, results in non-equidistant sampling of the k-space in the frequency-encoding direction and leads to artifacts, which manifest as stretching of the imaged objects in the frequency-encoding direction [Jones2010, Chap. 12, pp. 197-198].

Another common feature of EPI is the back-and-forth trajectory of k-space sampling. The k-space lines are sampled in the +X (forward) direction for one line (odd), and -X (reverse) for the other line (even) etc. At the same time, there are small, but non-negligible delays between the gradient waveforms and the sampling window. This results in the misalignment of k-space lines with respect to each other. The consequence is the appearance of Nyquist ghosts after image reconstruction – improperly localized intensity components partially obscuring the object [Jones2010, Chap. 12, pp. 184-185].

Both ramp sampling and ghosting artifacts need to be corrected during image reconstruction from EPI k-space data. Re-gridding is performed to resample non-equidistant samples into a Cartesian grid. The resampling matrix Q is data-independent and is constructed using the sinc interpolator [Jones2010, Chap. 12, pp. 197-198]:

$$Q(p,r) = \frac{\sin(k_x(p) - k_{\hat{x}}(r))}{k_x(p) - k_{\hat{x}}(r)} \qquad \text{Equation 3}$$

where $k_{\hat{x}}$ are coordinates of points in the desired Cartesian grid, $k_x$ are the coordinates of acquired points of a k-space line defined by the integral of the gradient field $G_x$ (Equation 2), and p and r are the resampling matrix coordinates, both matching the number of points in the k-space line $K_x$. Matrix Q is applied by a dot product to each ramp-sampled k-space line:

$$K_x^{regridded} = K_x^{ramp-sampled} \cdot Q \qquad \text{Equation 4}$$

The re-gridded k-space is Cartesian, but it is still affected by Nyquist ghosting, which is corrected in the consecutive step. One of the common methods requires acquisition of additional k-space lines, called navigators, passing through the centre of k-space. These lines are acquired in the opposite (+X and -X) directions, which allows to compute their phase differences. This is performed in the hybrid x-$k_y$ space, after the inverse Fourier transform of the navigator lines $N_1$, $N_2$ and $N_3$ along X. The linear phase shift is estimated as [Hansen]:

$$\theta_{ghost} = \measuredangle \left\{ \sqrt{(\mathcal{F}_x^{-1}(N_1) + \mathcal{F}_x^{-1}(N_3))^* \circ \mathcal{F}_x^{-1}(N_2)} \right\} \qquad \text{Equation 5}$$

where $\mathcal{F}_x^{-1}(N_1)$, $\mathcal{F}_x^{-1}(N_2)$ and $\mathcal{F}_x^{-1}(N_3)$ are inverse-Fourier-transformed navigator lines with $\mathcal{F}_x^{-1}(N_1)$ and $\mathcal{F}_x^{-1}(N_3)$ being acquired in the same direction in k-space (e.g. forward, +X) and $\mathcal{F}_x^{-1}(N_2)$ being acquired in the opposite direction (e.g. reverse, -X), ∘ denotes a sample-wise product, * denotes the complex conjugate. The phase shift is applied to re-gridded lines in x-$k_y$ space as:

$$\mathcal{F}_x^{-1}(K_x^{corrected}) = \mathcal{F}_x^{-1}(K_x^{regridded}) \circ e^{j\theta_{ghost}}, \text{ for all lines acquired in the forward direction (Equation 6),}$$

$$\mathcal{F}_x^{-1}(K_x^{corrected}) = \mathcal{F}_x^{-1}(K_x^{regridded}) \circ (e^{j\theta_{ghost}})^*, \text{ for all lines acquired in the reverse direction (Equation 7),}$$

yielding ghosting- and ramp-sampling-corrected x-$k_y$ space lines $\mathcal{F}_x^{-1}(K_x^{corrected})$. The data in the corrected x-$k_y$ space are then Fourier-transformed along Y to obtain the reconstructed image.

The described reconstruction algorithm is implemented in the 'ismrmrd-python-tools' toolbox authored by David Ch. Hansen [Hansen].

## 3.2. Simulation of brain MRI

With imaging parameters typical for T2-weighted imaging, we used the BrainWeb phantom and the spin echo DWI signal equation to calculate a b=0 DWI image:

$$I = k \cdot \rho \cdot e^{-\frac{TE}{T_2}} \left(1 - e^{-\frac{TR}{T_1}}\right) e^{-bD} \qquad \text{Equation 8}$$

where I is the signal intensity, k is a gain factor, $\rho$ is the proton density, $T_1$ and $T_2$ are tissue relaxation times, TE and TR are imaging parameters, b is the b-value and D is the apparent diffusion coefficient. b was set to 0, since at this point the BrainWeb phantom does not include reference ADC values for the tissues.

I is calculated for each voxel of the phantom and for each tissue contributing to the voxel volume. The result is a realistic modulus brain image $I_m$ with a relatively high isotropic resolution and devoid of any artifacts such as noise or blurring, besides the effects of mixing multiple tissues within a voxel (partial volume effect).

## 3.3. Modelling of phase maps and introducing phase variation

To obtain complex images from the modulus BrainWeb image $I_m$, we apply phase variation. This is achieved by multiplying $I_m$ with the complex phase term:

$$I_c = I_m \, e^{j\theta}, \qquad \text{Equation 9}$$

which yields a complex-valued slice $I_c$ with non-zero real and imaginary parts. Equation 9 is applied to each slice of a 3D image, taking into account that phase variations are slice-specific. Phase variation is determined by phase maps $\theta$. We derived those from acquired brain DWI using bivariate polynomial fitting.

First, real-MRI phase maps were computed from complex-valued DWI, and phase-unwrapping was applied. Then, a bicubic fit was applied. Phase outside the brain region is theoretically null (in practice it shows variations arising from the noise). Due to that, the error of fit computation was limited to the brain region. The latter was obtained by DWI segmentation using simple thresholding and morphological operations.

Limiting the scope of a nonlinear fit to the brain region results in non-realistically rapid phase variations outside the brain region. For this reason, phase map values outside the brain region were obtained from the bilinear fit applied identically as described above, while from the bicubic fit we only selected phase values for the brain region. Because this disrupted smoothness of phase variation at the edges of the brain and background regions, we applied 5-fold Gaussian blurring with a small sigma value (0.75).

Figure 1 shows the real and imaginary components of a coronal slice of the BrainWeb after introducing phase variation. While $I_m$ was positive-valued, the components of $I_c$ contain negative values as well, with image background remaining at the value of 0.

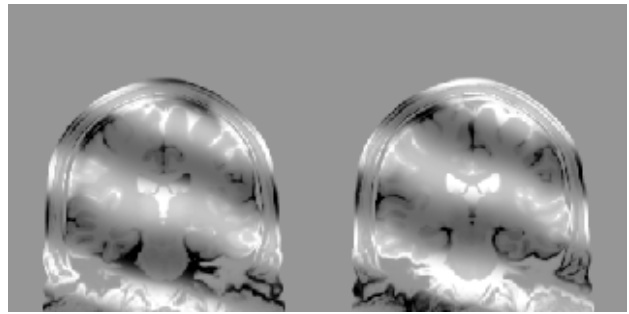

**Figure 1. Real (left) and imaginary (right) components of a coronal slice of a b=0 image simulated using Equations 8, 9 and the BrainWeb phantom**

### 3.4. Modelling and simulation of EPI effects

To model EPI data we reverted the EPI reconstruction and correction pipeline described before and applied it to the complex BrainWeb image $I_c$. In the reverted pipeline, EPI effects are introduced in the following order: Nyquist ghosting, ramp sampling effects and noise.

To introduce Nyquist ghosting, complex images are Fourier-transformed along Y to the x-$k_y$ space. From Equation 5, we obtained the phase correction values. These are not directly applicable to other images such as the BrainWeb, because phase, in the X direction, varies only within the extent of the tissue.

The phase differences causing Nyquist ghosting are linear [Jones2010, Chap. 12, pp. 190-191, EPIbook]. In the calculated $\theta_{ghost}$, phase correction values are readily normalized to the range of [-1, 1]. We utilize this property and construct reverse phase shift vectors $\widehat{\theta_{ghost}}$ in the value range of [1, -1]. This linear phase variation extends only within these voxels which contain tissue, as mentioned. To find those in each image row, we used a segmentation of the BrainWeb image, obtained by simple thresholding, taking into account that background voxels in this image are zero-valued by design. Phase de-correction vectors are applied as:

$$[\mathcal{F}_y(I_c)]_x^{ghosting} = [\mathcal{F}_y(I_c)]_x \circ e^{j\widehat{\theta_{ghost}}}, \text{ for odd rows} \qquad \text{(Equation 10)}$$

$$[\mathcal{F}_y(I_c)]_x^{ghosting} = [\mathcal{F}_y(I_c)]_x \circ \left(e^{j\widehat{\theta_{ghost}}}\right)^*, \text{ for even rows} \qquad \text{(Equation 11)}$$

Figure 2a shows a slice from the image with introduced Nyquist ghosting, obtained by the inverse Fourier transform along Y of all rows $[\mathcal{F}_y(I_c)]^{ghosting}$ in the x-$k_y$ space.

The resulting x-$k_y$ space data with simulated linear phase shift are transformed to the $k_x$-$k_y$ space (k-space), in which the inverse re-gridding operation is applied.

We computed the re-gridding kernel $Q$ using Equation 3 and the sampling parameters of the acquired brain DWI images. In order to introduce ramp sampling effects, we reverted the effect of re-gridding by applying an inverse kernel R, obtained via the Moore-Penrose pseudoinverse. R was applied to each line of k-space (Figure 2b) as:

$$K_x^{ghosting, \ ramp-sampling} = [\mathcal{F}_{xy}(I_c^{ghosting})]_x \cdot R \qquad \text{(Equation 12)}$$

Finally, noise was added to the image with ghosting and ramp sampling artifacts (Figure 2c). We used realistic noise maps, acquired via the air scan. Taking advantage of the linearity of the Fourier Transform, noise addition was performed in the image domain instead of the k-space. Image domain noise was obtained from the air scan's k-space data with the same reconstruction method as for the case of patient images.

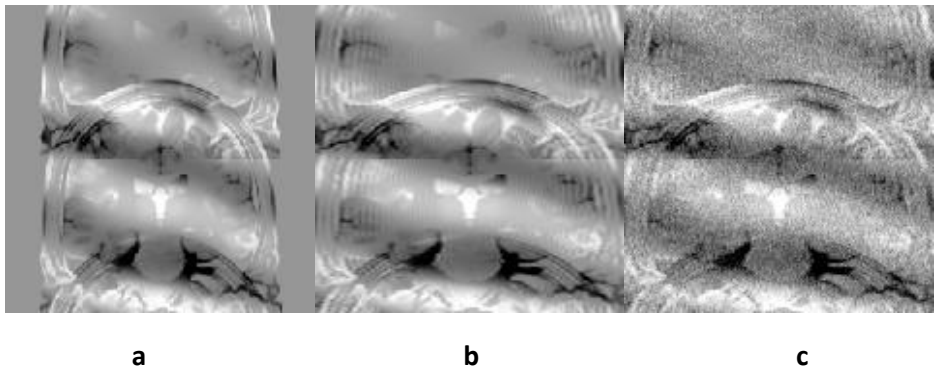

**a**          **b**          **c**

**Figure 2.** Real component of the simulated coronal slice. Image with introduced ghosting (a), image with ghosting and stretch resulting from ramp sampling (b), image with ghosting, stretch and additive Gaussian noise (c)

### 3.5. Resampling of the modulus BrainWeb volume

Phase maps', noise maps', ghost correction kernels' and regridding kernels' dimensionality is directly dependent on image resolution. We derived them from the acquired brain DWI, which had 160x320x25 voxel resolution. Hence, the modulus BrainWeb image was resampled to match this resolution, but this was achieved in two ways, differently for the training and test data, mainly due to technical reasons.

To create the training dataset, we chose 25 coronal slices. These slices retained their original thickness of 1 mm. We found this relevant after pilot experiments, which showed that the network tends to remove fine details and better results can be obtained when the training data contain such details. Then, we resized the slices from 181x217 voxels to 125x125, first by cropping to 180x180, and then by downsampling with nearest-neighbour interpolation. After this resampling step, the spatial resolution of the slices in the slice plane was closely matching the resolution of the acquired brain DWI. 25 125x125 slices where then embedded in the zero-filled array with 160x320x25 voxels. The brain in coronal slices occupies the lower part of the slices (Figure 2). For this reason and because the phase variation maps were obtained for axial slices, with the most accurate bicubic fit in the central region of the image, the coronal slices were shifted in order to move the brain to the image centre. After introducing phase variation, they were shifted back to the original position without the need for interpolation.

For testing of the neural network, where we wanted to achieve the greatest possible similarity of the test images to the real ones, we used axial slices of the BrainWeb image. In this case, the whole volume was affinely registered to the brain b=0 DWI volume of a chosen patient and linearly resampled to match the resolution. This readily resulted in a 160x320x25 voxel real-MR-registered BrainWeb volume.

### 3.6. Data preparation for learning and evaluation

The training and validation sets were prepared from resampled coronal BrainWeb slices. In the pilot experiments, we applied the complete EPI data model, i.e. phase variation, ramp sampling artifacts, ghosting artifacts and noise. The networks trained on the images from the complete EPI data model, however, tended to partially remove object edges alongside with noise. We interpreted this effect as the result of lowered training image resolution due to the introduced ghosting and ramp sampling artifacts. Network trained on images without these artifacts lead to better results and we finally decided to only introduce phase variation and noise to training and validation images. To increase the number of training examples, we used three different noise maps for each of the 25 coronal slices. Selecting different noise maps from the total of 25 was achieved by random permutations without the possibility to repeat. Due to this the network has a possibility to observe the same object with three different random noise maps, but also to observe the same noise map obscuring three different objects, as it searches for the optimal weight values. At this stage, the prepared training data consisted of 75 complex-valued coronal slices. The complex components were then separated, which doubled the number of training examples. In the final stage of data preparation, we selected every 9$^{th}$ training example and moved it into the validation set.

Test data were prepared by applying the complete EPI data model (phase variation, ghosting, ramp sampling, noise) to axial BrainWeb slices, previously registered to one of the real brain DWI images. To guarantee fair testing, we used different phase and noise maps for the test slices to be denoised by the neural network.

### 3.7. Modelling patient motion between repeated scans

Patient motion during an imaging experiment is practically inevitable and, due to blurring and other artifacts, influences the effective image resolution. This happens because of involuntary rigid motion, but also due to motion of internal organs – bowel motion, breathing-induced motion, muscle contractions etc. In the experiments, we use a motion model derived from the acquired brain DWI.

Thirty-two repeated b=0 volumes were affinely registered using mean squared error as the loss function. The resulting affine matrices for each volume were used to introduce inter-scan motion in the test images (axial real-MR-registered BrainWeb slices). Zeroth order interpolation was used to find voxel intensities in the transformed grid and to keep sharp edges. In the case of moving images, noise was added after introducing the affine transformation.

## 3.8. Modelling the Point Spread Function

The Point Spread Function (PSF) is the effective blurring function that affects imaging systems. In MRI, many phenomena lead to blurring. Based on the central limit theorem, we assume that the MRI scanner's PSF is approximately Gaussian.

To estimate the PSF, we use the acquired images of an AJR-approved phantom. The phantom has a large-diameter cylindrical shape and is placed with the cylinder base approximately in the axial plane. We study the intensity profiles drawn between the background and the phantom, normally to the latter. As described in [Materka, 1991], the sigma parameter of a Gaussian PSF can be estimated from such a profile by fitting the Gaussian error function to the measured intensity profile (Figure 3).

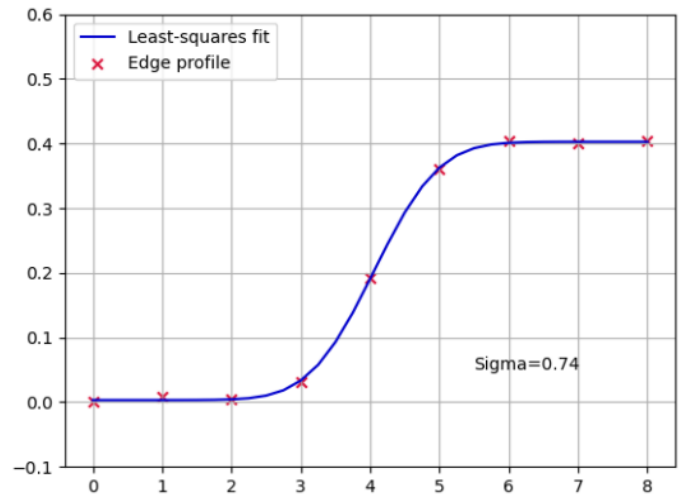

**Figure 3. Intensity profile (red crosses) with the fitted model (blue line), taken from the phantom image through the border of the phantom and the background, normally to the phantom curvature. Sigma was estimated to 0.74 in this example**

We examined four slices from the central region of the phantom. We found the standard deviation of the Gaussian PSF to be in the range of 0.61-0.75, depending on the slice. In the experiments, we adapted the value of 0.65.

## 3.9. Modelling of noisy images with different SNR

In our experiments, we defined the image SNR as the ratio of the mean value of the noise-free object in the modulus domain (including air regions like sinuses, but excluding background) to the standard deviation of the noise in the complex image domain. For simplicity, we set the noise standard deviation to 1 in all noise maps (details of this procedure are explained in the Section 3.10.). Then, images with different SNR are obtained by dividing the modulus image by its mean value and then multiplying by the value of the desired SNR. Afterwards, the unit standard deviation noise is added. Further in the experiments, we consider SNR values between 1-9.

## 3.10. Measurement and analysis of real MRI noise

All images in our study were normalized to have the same noise level, independent of the SNR. For this purpose, we developed a method for measurement of the noise statistics in complex DWI images of the brain.

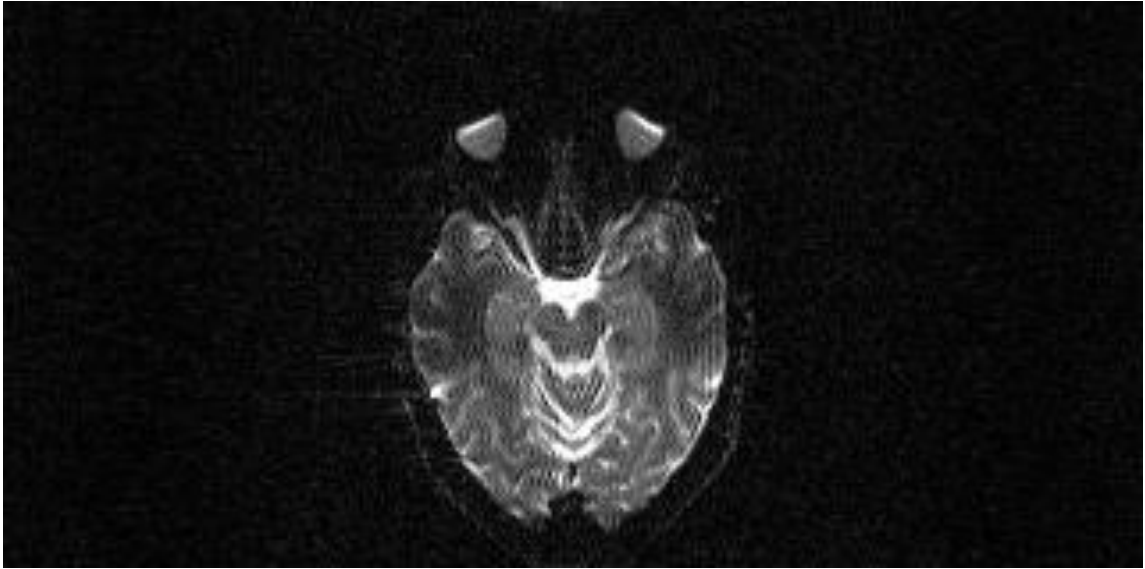

**Figure 4. Modulus b=0 image of the brain. The frequency-encoding direction is oversampled by a factor of 2, which results in acquisition of a wide background region that can be used to normalize different images**

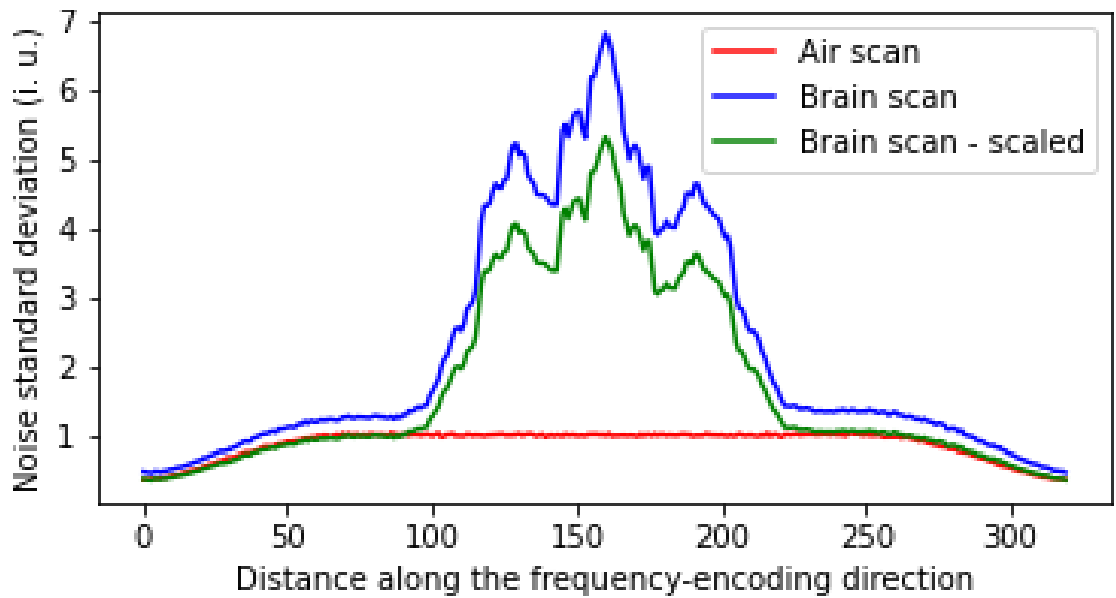

**Figure 5. Noise standard deviation profiles drawn for one of the coils in the air scan (red) and the same coil in the brain scan (blue). Standard deviation of noise is elevated in the latter due to the presence of a patient in the scanner. Proper scaling of the brain image intensities normalizes the noise levels (green). The scaling factor is chosen such that it minimizes the average error of the noise profiles in the oversampled region (0-79, 240-319 on the horizontal axis of the plot)**

MRI images are often acquired with oversampling in the frequency-encoding direction of the k-space. This is primarily performed to avoid aliasing in this direction. As a side effect, the field of view of the image is extended proportionally to the oversampling factor. In the case of the acquired brain DWI (Figure 4), the oversampled regions contain noise only. We observe that image domain noise is non-stationary in the oversampled regions, while it is roughly stationary the non-oversampled region. In particular, the noise standard deviation changes along the frequency-encoding direction. When plotted, this statistic forms characteristic profiles, as shown in Figure 5 (red line), in accordance with [Kellman, 2005]. The profiles are conserved up to their shape in images acquired with different b-value, coils or resolution, but vary in scale. From this property, we draw a conclusion that it is possible to normalize real brain DWI images according to the noise level by matching noise standard deviation profiles in the background region.

To standardize the images in our study, we first normalized the air scans, containing noise only, such that they had on average unit standard deviation within the central, non-oversampled region of the noise profile (Figure 6, red line). Then, for patient images, we found scaling factors that minimized the error between their noise profiles and the noise profiles of the normalized air scans. The error was computed only in the background region, which, in the case of the acquired brain images, was assumed to be the oversampled region. The scaling factors were then used to normalize brain images (Figure 6, blue and green lines).

### 3.11. Neural network architecture for DWI denoising

Based on our previous experience, we use a simple but efficient convolutional neural network model – the super-resolution convolutional neural network (SRCNN). This network is composed of three convolutional layers and their mathematical function was described in detail in [Dong2016, Jurek2020]. For the denoising purpose, we adapt the SRCNN similarly as in [Zhang2017]. This is, we force the network to estimate the noise map instead of the denoised image (Figure 6). In this setting, the network acts as a noise-pass filter, which we call the CNN filter from this point. The CNN-estimated noise map is simply subtracted from the input image, yielding a denoised output. The denoised output is compared to a clean reference and the loss function is evaluated. We used the mean squared error as the final loss function of choice, though we also experimented with L1 loss and KL divergence.

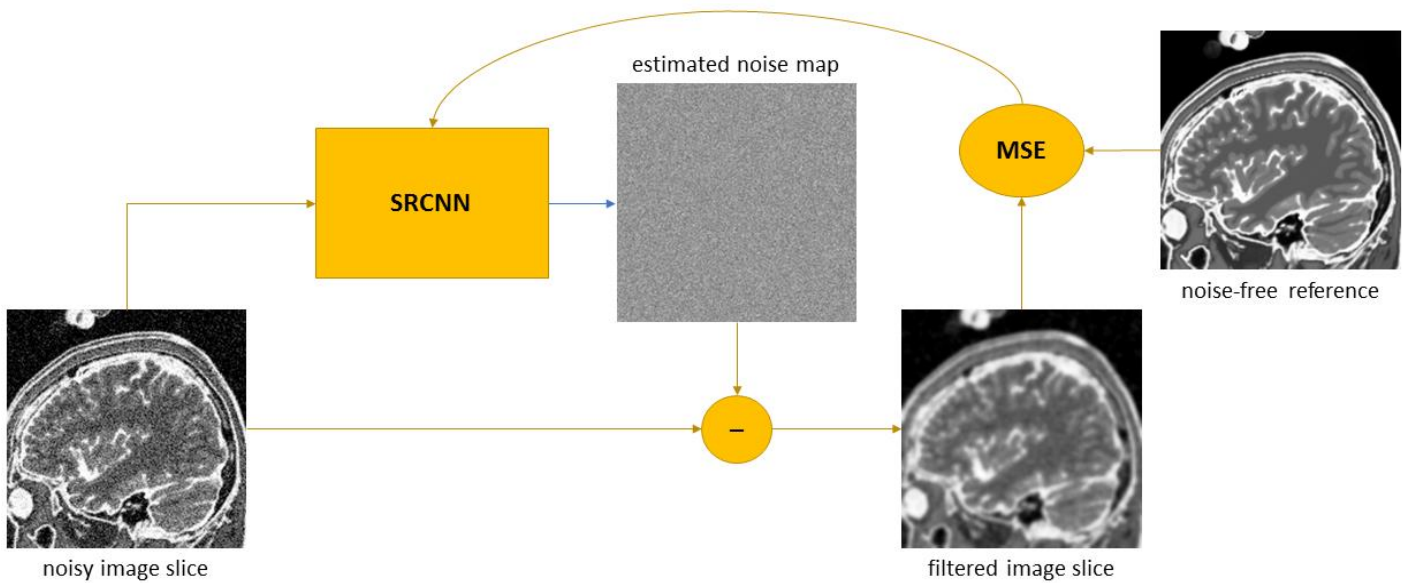

Figure 6. Scheme of the denoising algorithm using the SRCNN.

### 3.12. Neural network training

Training of the neural network was accomplished with Adam optimizer, minimizing the mean squared error between the denoised images and a noise-free reference. The learning rate was set to 0.001. Optimization was performed on mini-batches of 9 randomly shuffled training examples within each epoch of training. We allowed the network to learn for 10000 epochs, unless the early stopping mechanism detected overfitting or lack of significant improvement of the loss function evaluated on the validation set. The early stopping algorithm was the same as described in [Jurek, 2020], with patience set to 50 and **minimal improvement** of $10^{-8}$. When training stopped, model weights with the lowest validation error were selected from training history.

## 4. Experimental results

### 4.1. Studying the SSIM and PSNR of denoised artificial images

We evaluated CNN-denoised BrainWeb slices in reference to clean ground truth slices. Denoised images were compared to the average of repeated images using the SSIM and PSNR. Averaged images were motion-free and were obtained using two approaches – averaging in the complex domain and averaging in the modulus domain.

We considered two test data models – the first one represented images with a simulated PSF. The other model did not include any PSF. We considered SNR values of 1, 3, 5, 7 and 9.

Denoising was applied in two ways – either before Nyquist ghosting correction and regridding or afterwards. We refer to these options as pre-correction and post-correction denoising.

### 4.1.1. EPI model with the Point Spread Function applied

Including the PSF in the model better suits the test data to DWI images, that are typically more blurry than conventional spin-echo images. We applied in-plane blurring of the slice before noise addition. We selected the sigma value of the Gaussian filter to be 0.65 for each slice, as mentioned before.

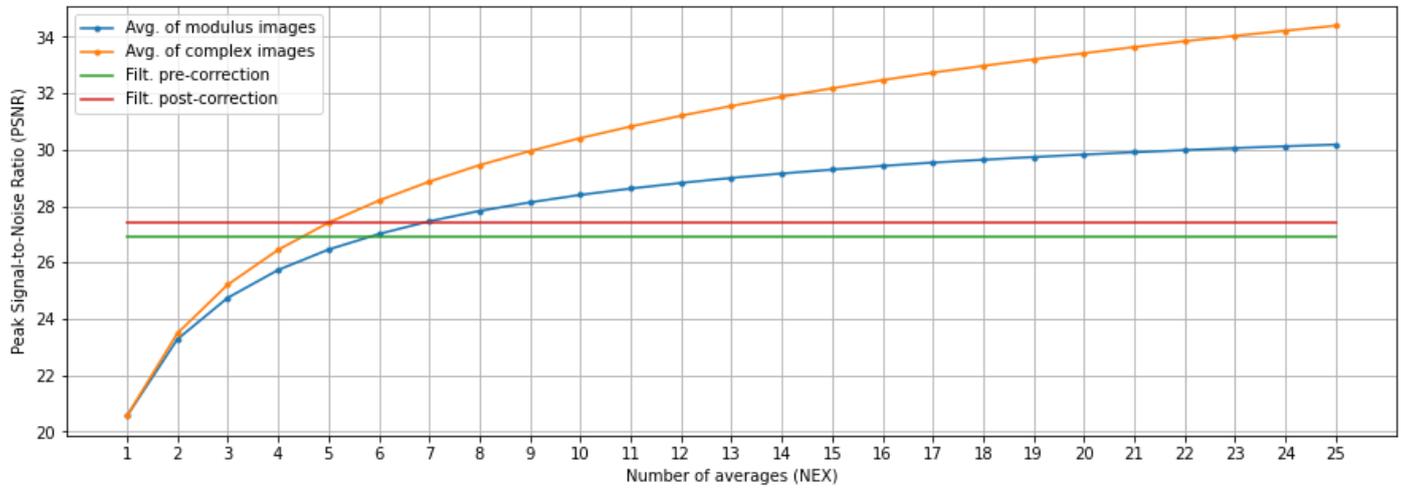

**Figure 7. PSNR plot versus the NEX for the averaged images with SNR = 3. Pre- and post-correction denoised NEX = 1 image PSNR values (for tissue) are shown as horizontal lines. Model with PSF and no motion.**

Quantitative results are analysed first. Figure 7 shows PSNR and SSIM values obtained by complex and modulus domain averaging, as well as pre- and post-correction denoising with a CNN. Here, the case of SNR = 3 is presented, and the remaining plots for other SNR values are published in the Appendix (Figures A1-A8).

We notice that averaging of repeated images, both in the modulus and complex domains, brings improvement of the metrics with respect to the clean ground truth image. As expected, complex domain averaging is superior to modulus domain averaging, especially for the lower SNR values and in terms of the PSNR, but it of limited utility in DWI imaging due to random phase variations in complex images.

Denoising single-NEX images, both pre- and post-correction, improves the quality measured by the SSIM and PSNR. Pre-correction denoising consistently yields lower metric values than post-correction denoising. The difference between the resulting denoised images is shown in Figure 8. As visible, denoising the complex images before ghosting correction leads to errors and ghosts are not fully removed.

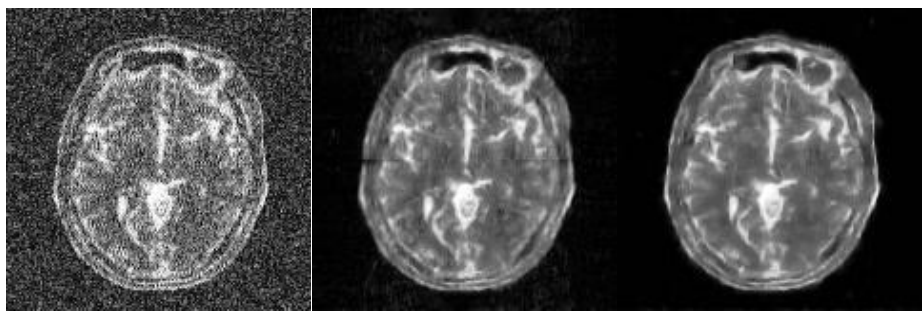

**Figure 8. From left to right: noisy image, pre-correction denoising, post-correction denoising. SNR = 3.**

Instead of comparing individual values of the PSNR and SSIM, we rather examine the number of averaged scans necessary to approximately match the denoising result, according to the SSIM and PSNR. We read off these numbers from the PSNR vs. NEX and SSIM vs. NEX plots, as presented in Table 1. They are presented separately with respect to averaging in modulus and complex domains. We also consider comparison of full slices (including noisy) background and comparison of tissue voxels only, which is achieved after zeroing the background in denoised, averaged and clean images.

|  | MODULUS AVERAGING | | | | COMPLEX AVERAGING | | | |
|---|---|---|---|---|---|---|---|---|
|  | SSIM (slice) | SSIM (tissue) | PSNR (slice) | PSNR (tissue) | SSIM (slice) | SSIM (tissue) | PSNR (slice) | PSNR (tissue) |
| SNR = 1 | >25 | 16 | >25 | >25 | >25 | 8 | 25 | 12 |
| SNR = 3 | >25 | 7 | >25 | 7 | >25 | 6 | 17 | 5 |
| SNR = 5 | >25 | 6 | >25 | 4 | >25 | 6 | 13 | 4 |
| SNR = 7 | >25 | 5 | >25 | 3 | >25 | 5 | 10 | 3 |
| SNR = 9 | >25 | 5 | >25 | 3 | >25 | 5 | 10 | 3 |

**The necessary number of repeated scans to obtain metric values (PSNR, SSIM) close to the CNN-denoising result. Averaging of motionless scans with the PSF.**

The necessary numbers of repeated scans, as shown in Table 1, are diverse. First, they depend on the SNR of the noisy image. Consistently, the efficiency of CNN-denoising drops with a growth in SNR compared to averaging. For example, at the SNR of 3, 7 repeated scans would be necessary to reach a similar PSNR value to CNN-denoising, but only 3 for the SNR of 9.

Secondly, we observe that including image background in the comparison significantly favours CNN-denoising. As far as background itself is not a medically relevant region, denoising in the background might be of value for head segmentation, for example.

Thirdly, we note that SSIM and PSNR tend to give divergent results. While only 3 repeated scans would match the PSNR of a CNN-denoised image at SNR=9, to reach a similar SSIM 5 scan repetitions would be necessary. In general, we observe that more repeated scans are necessary to match the SSIM value than the PSNR.

Following quantitative evaluation, we perform a visual exam of the denoising quality (Figure 9). In order to complete this step, we compare denoised images with modulus averaged images, where the NEX of averaged images is chosen according to Table 1.

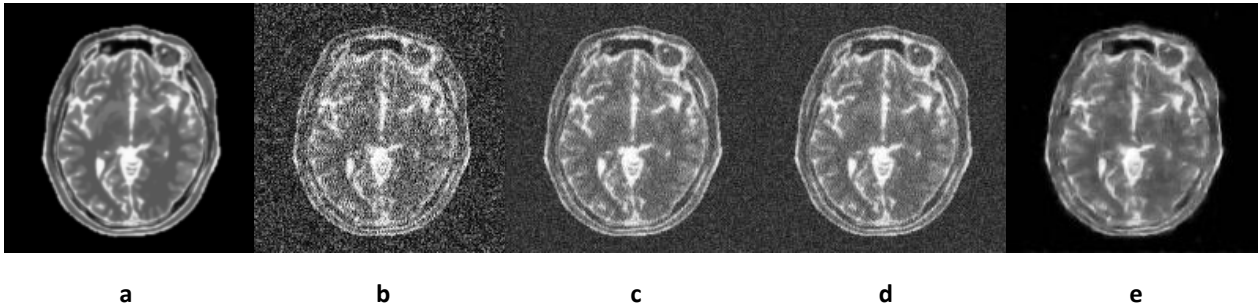

      a                    b                    c                    d                    e

**Figure 9. (a) clean slice, (b) noisy slice with SNR = 3, (c) ARI with NEX=6, (d) ARI with NEX=7, (e) CNN denoised image. Averages computed in the modulus domain (see also Appendix Figure A9).**

The averaged repeated images (ARI) and denoised images in Figure 9c, d, e are equivalent in terms of SSIM and PSNR, within the tissue. It is straightforward that the CNN significantly reduced the noise, so that it is barely noticeable in the denoised image. In contrast, averaged images still appear noisy both in the background and the tissue. However, the denoised image contains certain distortion that overwhelms the benefit of reducing noise. We analyse this problem in more detail in Section 4.3.

### 4.1.2. EPI model without the Point Spread Function

Although PSF effects occur in real-world MR images, high-resolution images are affected to a lesser degree. In such case, the image spectrum contains more high-frequency components. We performed a similar analysis with simulated images without the PSF (or sigma = 0). Table 2 shows NEX values necessary to reach the quality of denoised images as measured by the SSIM and PSNR.

|  | MODULUS AVERAGING | | | | COMPLEX AVERAGING | | | |
|---|---|---|---|---|---|---|---|---|
|  | SSIM (slice) | SSIM (tissue) | PSNR (slice) | PSNR (tissue) | SSIM (slice) | SSIM (tissue) | PSNR (slice) | PSNR (tissue) |
| SNR = 1 | >25 | 6 | >25 | >25 | 23 | 4 | 25 | 7 |
| SNR = 3 | >25 | 5 | >25 | 4 | >25 | 4 | 11 | 3 |
| SNR = 5 | >25 | 5 | >25 | 3 | >25 | 5 | 9 | 3 |
| SNR = 7 | >25 | 4 | >25 | 2 | >25 | 4 | 7 | 2 |
| SNR = 9 | >25 | 4 | >25 | 2 | >25 | 4 | 7 | 2 |

**Table 2. The necessary number of repeated scans to obtain metric values (PSNR, SSIM) close to the CNN-denoising result. Averaging of motionless scans without the PSF. For SSIM and PSNR plots, see Appendix Figures A10-A17)**

We notice that the denoising task in images without PSF blurring is more difficult, since the necessary NEX numbers are lower than the corresponding values in Table 2, but the remarks concerning the dependence of these numbers on the original SNR, the chosen metric and averaging mode generally hold. In this case, we noticed that at least the NEX of 2 was necessary for SNR=7 and 9 to match the PSNR of a denoised image within the tissue.

Figure 10 allows for a visual comparison of the clean, denoised, and averaged images.

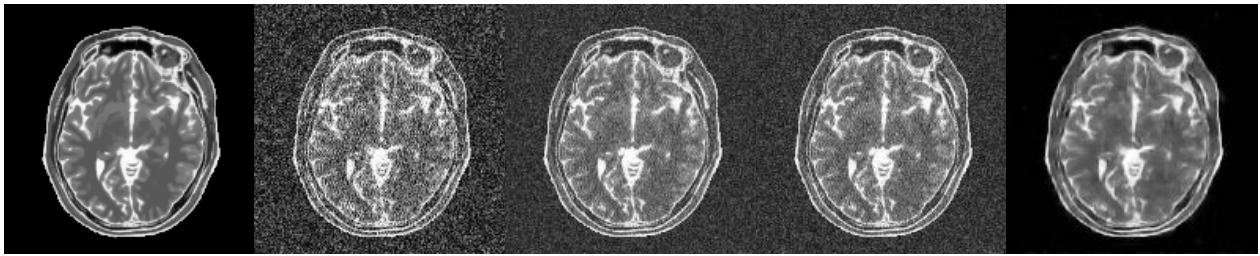

**Figure 10. From the left: clean slice, noisy slice with SNR = 3, ARI with NEX=5, ARI with NEX=4, CNN denoised image. Averages computed in the modulus domain.**

Similarly as in the case without the PSF, the denoised image is visually significantly less noisy than averaged images with a similar SSIM or PSNR value within the tissue. The NEX numbers providing similar metric values are lower than for the case with PSF, as mentioned. We attribute this effect to the presence of finer details and more sharp edges in the image, that are either more difficult to denoise or are recognized as noise and removed. This is further analysed in Section 4.3.

### 4.1.3. Model with motion

The previously described experiments neglected the fact of patient movement between the repeated scans. We include the latter effect in the next experiment. We tested the model with motion and without PSF blurring. Prior to averaging of repeated images, they were registered back using the inverse of the affine matrices used to introduce the movement. Bicubic interpolation was used for resampling. This procedure is in accordance with algorithms implemented in MR scanners, where registration of repeated images is performed in the image domain as a postprocessing step. Again, we calculated SSIM and PSNR values of averaged images and the denoised ones, but we only considered post-correction denoising this time. Figure 11 shows the PSNR plot obtained for SNR=3 in the tissue region (for other SNR values and SSIM plots, see Appendix Figures A18-A27).

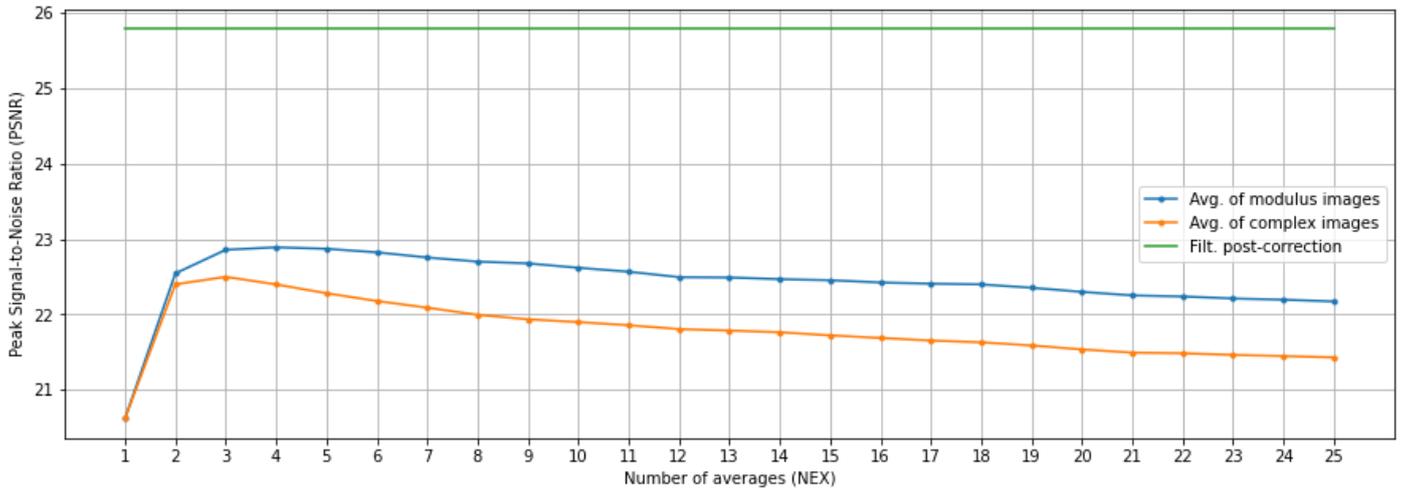

**Figure 11. PSNR plot versus the NEX for the averaged images with SNR = 3. Post-correction denoised NEX = 1 image PSNR value (for tissue) is shown as a horizontal line. Model without PSF and with motion.**

From Figure 11 it is clear, that averaging moving volumes has limited ability to improve their similarity to a motionless and noise-free ground truth. For the case of SNR = 3 and the particular movement pattern derived from one of the real DW images, the maximum PSNR was obtained for NEX=4, but was still lower than the PSNR of motionless ARI with the same NEX (23 vs. 26 dB). For an unclear reason, averaging performed better in the modulus domain. This was observed for the PSNR in the tissue and for other SNR values, as well. Figure 12 shows a comparison of the clean, noisy, averaged and denoised images.

Finally, we note that the denoised images had significantly higher metric values in the tissue region that any of the ARI images for SNR>=5, as shown in Table 3.

|         | MODULUS AVERAGING | | | | COMPLEX AVERAGING | | | |
|---------|------|------|------|------|------|------|------|------|
|         | SSIM (slice) | SSIM (tissue) | PSNR (slice) | PSNR (tissue) | SSIM (slice) | SSIM (tissue) | PSNR (slice) | PSNR (tissue) |
| SNR = 1 | >25 | 5 | >25 | >25 | >25 | 4 | >25 | 10 |
| SNR = 3 | >25 | 8 | >25 | >25 | >25 | 8 | >25 | >25 |
| SNR = 5 | >25 | >25 | >25 | >25 | >25 | >25 | >25 | >25 |
| SNR = 7 | >25 | >25 | >25 | >25 | >25 | >25 | >25 | >25 |
| SNR = 9 | >25 | >25 | >25 | >25 | >25 | >25 | >25 | >25 |

**Table 3. The necessary number of repeated scans to obtain metric values (PSNR, SSIM) close to the CNN-denoising result. Averaging of motion-corrupted scans without the PSF.**

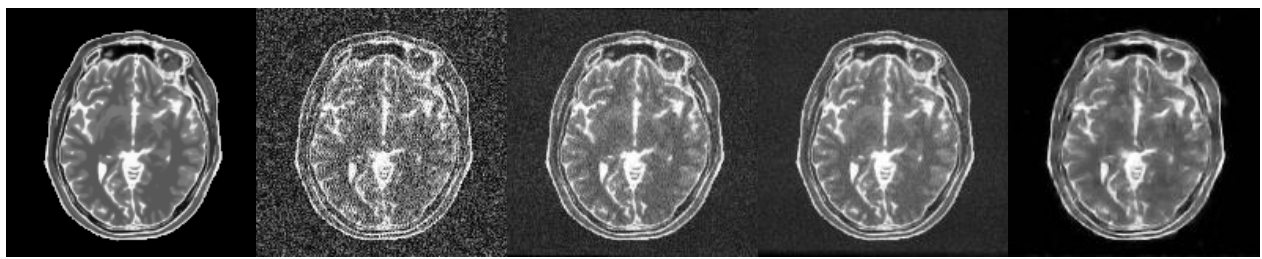

**Figure 12. Clean, noisy with SNR = 3, ARI with NEX=8 and motion, ARI with NEX=24 and motion, CNN-denoised image. Averages computed in the modulus domain (see also Appendix Figure A28).**

## 4.2. Application of the denoising neural networks to real DWI

Finally, neural networks trained using simulated images were used to denoise real head DWI images at b=0, 300 and 1000. We selected the networks trained for SNR=7, 5 and 3 for b=0, 300 and 1000, respectively. These SNR values roughly match the actual SNR values of the DWI images at these b-values.

First, we visually compared the results of pre- and post-correction denoising (Figure 13). Similarly as in the case of simulated images, ghosting correction is erroneous when denoising is applied before correction.

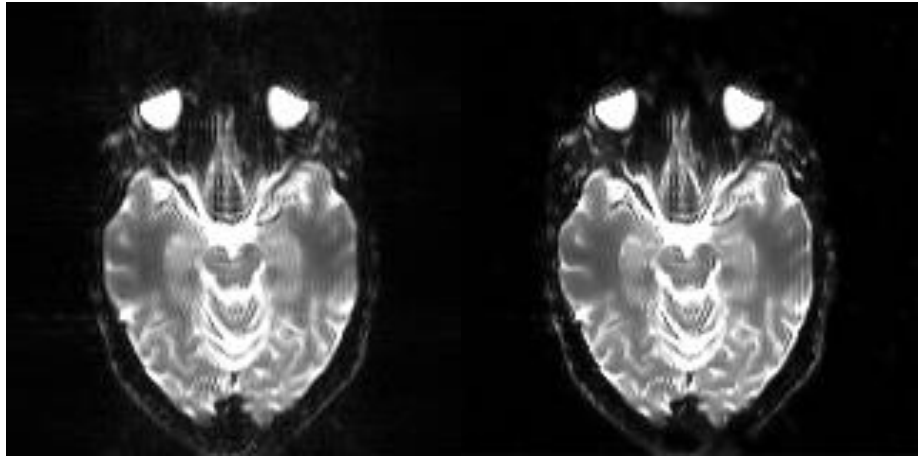

**Figure 13. Pre-correction denoising (left) leads to erroneous ghosting correction and blurriness as compared to post-correction denoising (right). A denoised slice from the b=0 DWI is presented.**

Therefore, we further present solely post-correction denoising results.

Denoising is performed after corrections in complex slices, separately for the real and imaginary components and for each coil and b-value. The case of one coil in the b=0 image is presented in Figure 14. More NEX values and other coils are shown in Figure A29 in the Appendix.

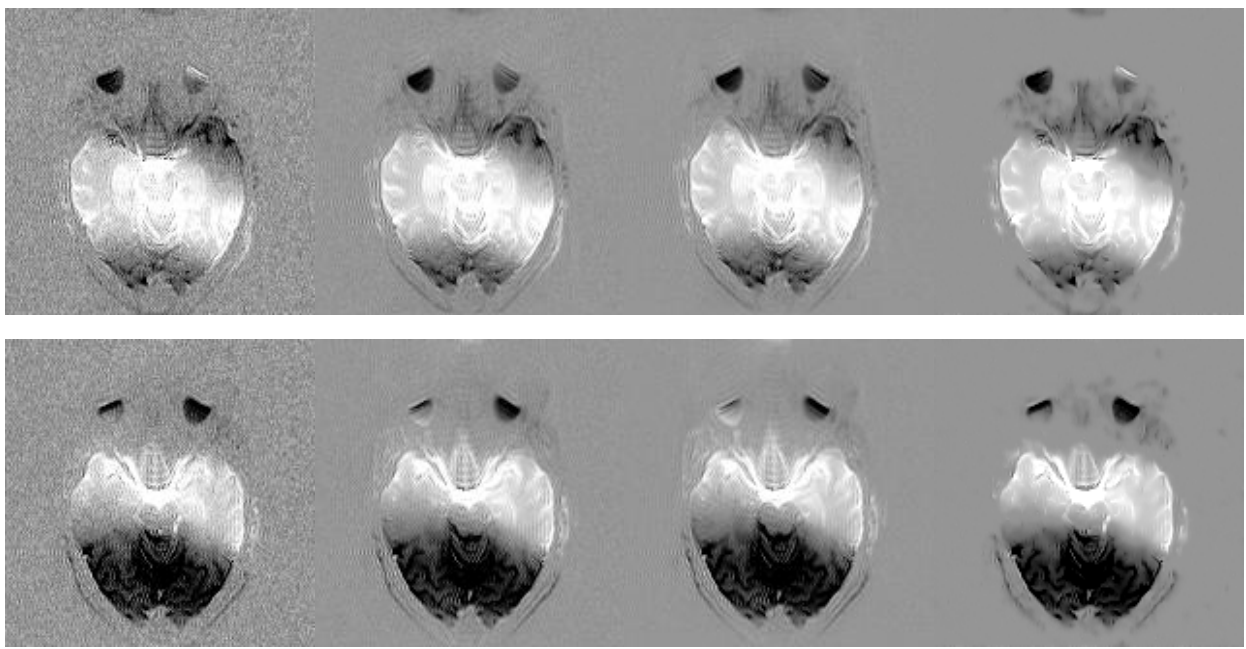

**Figure 14. Denoising results of the CNN trained with SNR=7 simulated data, compared to averaged images, for the case of the b=0 slice from receiver coil 1. From left to right: noisy image NEX=1, averaged NEX=9, averaged NEX=25, denoised image. Upper row: real part, lower row: imaginary part.**

Denoised complex components are combined to yield modulus images, for each coil. Then, slices reconstructed for each coil are combined using the square root of the sum of squares to yield the final multicoil slices. Figure 15 shows the multicoil slices obtained for different b-values (for the diffusion direction X in the case of b>0).

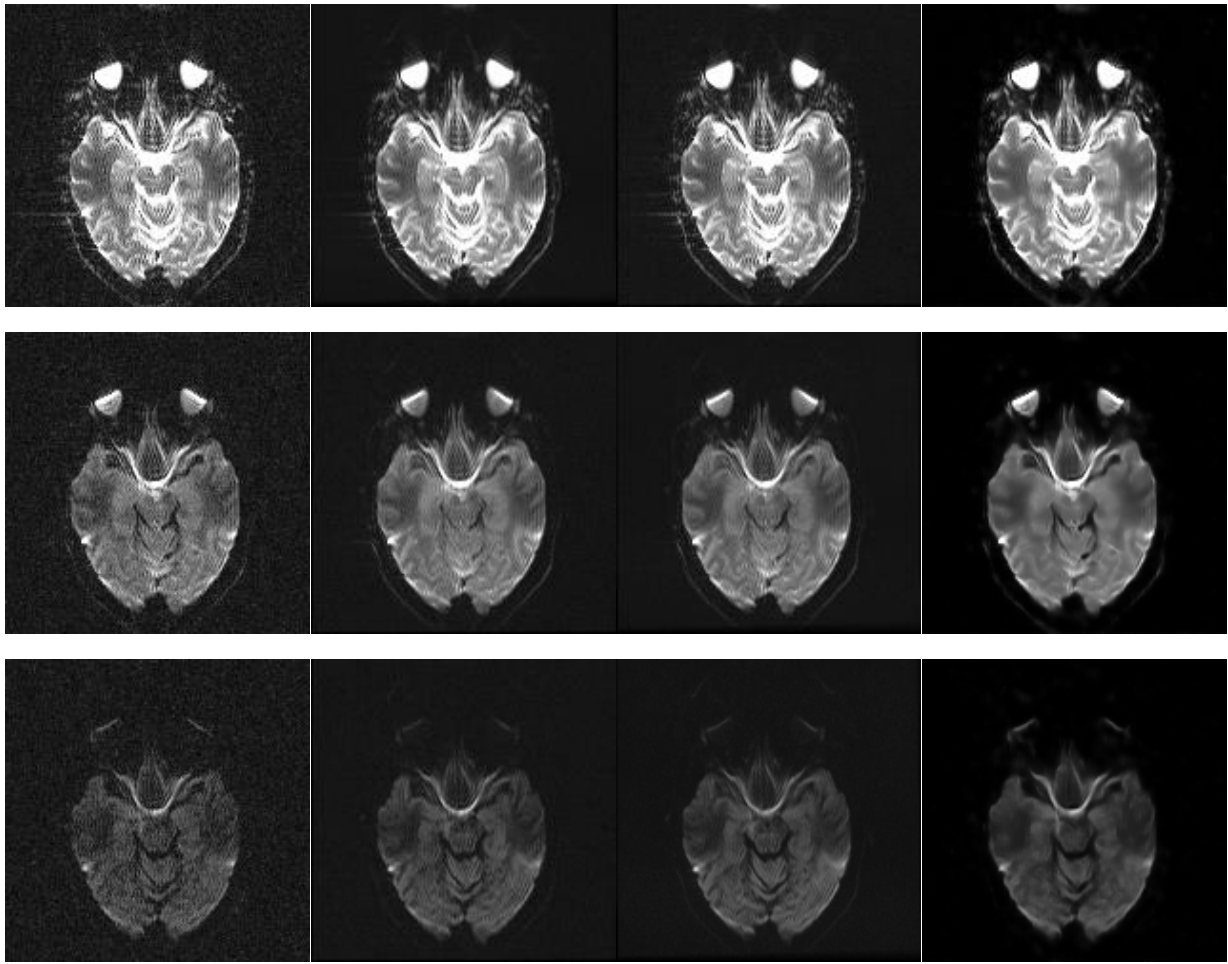

**Figure 15. Slices from b=0 (upper row), b=300x (middle row), b=1000x (lower row) DWI of the head. From left to right: noisy image NEX=1, averaged NEX=9, averaged NEX=25, CNN-denoised.**

MRI scanners usually do not export DWI in individual diffusion direction. Instead, diffusion trace images are computed by taking the geometric average of DWI intensities in the X, Y and Z direction (or more, if more advanced diffusion imaging is necessary). We then computed trace images for b=300 and b=1000, as shown in Figure 16.

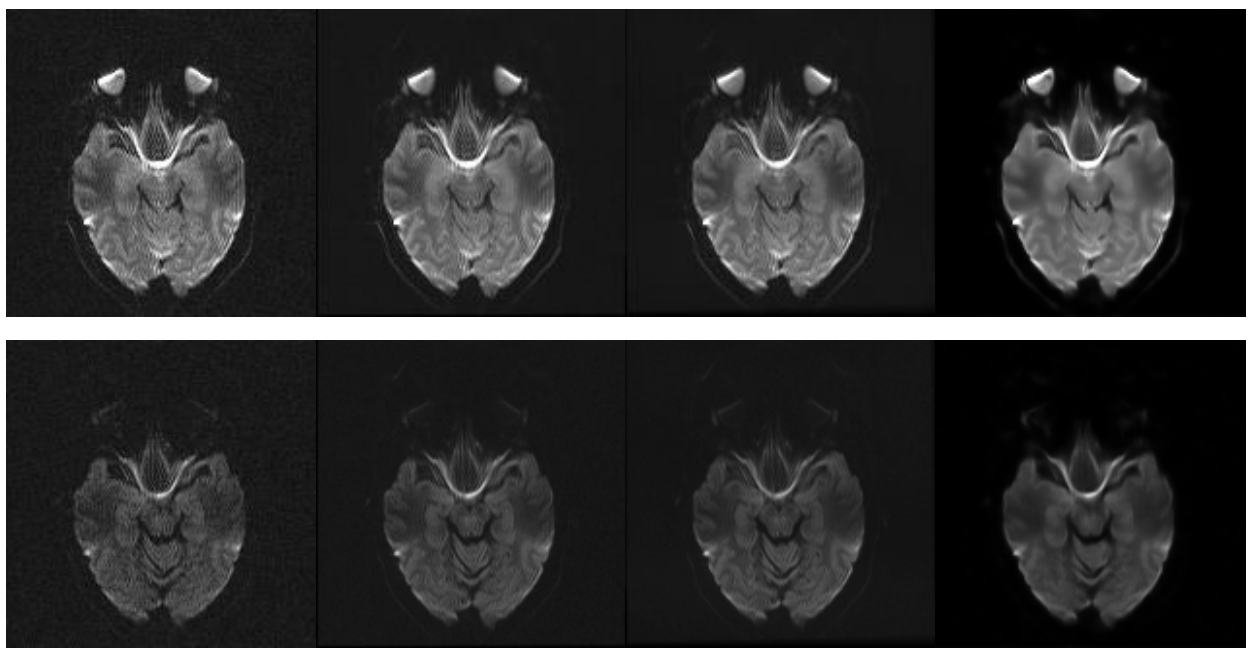

**Figure 16. Slices from b=300 trace (upper row) and b=1000 trace (lower row) DWI of the head. From left to right: noisy image NEX=1, averaged NEX=9, averaged NEX=25, CNN-denoised.**

Finally, the ADC maps are computed using the normal equation of linear regression using the b=0, b=300 and b=1000 trace images. ADC maps of one slice are shown in Figure 17.

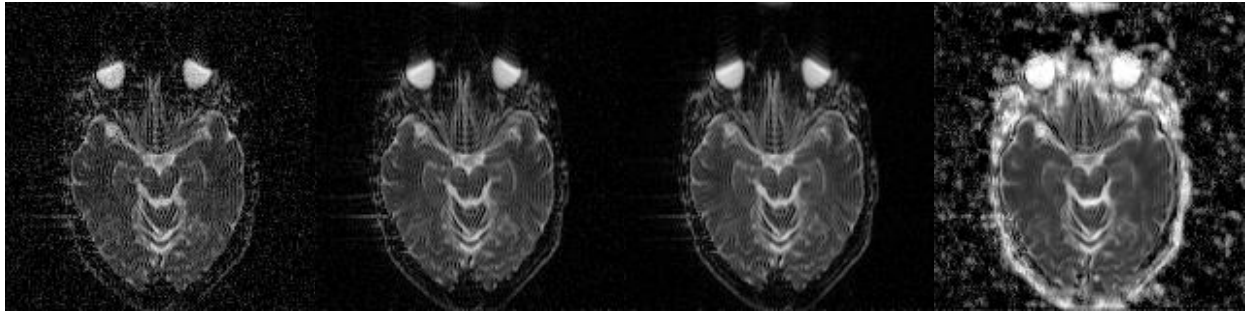

**Figure 17. Slices from ADC maps of the head. From left to right: noisy image NEX=1, averaged NEX=9, averaged NEX=25, CNN-denoised.**

When viewing denoised and averaged DWI images of the head, one can see that the neural network significantly reduced the noisy appearance of the image. Comparison of fine details shows that CNN-denoising did not lead to excessive blurring. This concerns both complex, modulus, multicoil and trace images.

In the ADC map computed from denoised images, we can see intensive artifacts, however, only outside the region of interest, i.e. the brain. These artifacts are manifested as high-ADC regions in the background, around the skull and also inside the eyeballs. Brain tissue seems not to be affected. We analyse these artifacts in Section 4.3.

### 4.3. Analysis of the denoising-induced image distortions

Despite observing improved visual, SSIM- or PSNR-measured quality of the denoised images compared to averaged ones up to certain NEX values, we perform a detailed analysis of the image content removed by the CNN filter.

First, we notice that the equivalence of SSIM or PSNR between denoised and averaged images, as shown in Figure 9, 10 and 12, does not imply that noise levels are similar. We can see that the NEX = 7 averaged image in Figure 9 is visually considerably more noisy than the denoised one. This implies that the CNN filter, despite reducing the noise, induces measurable distortion that deteriorates comparative metrics such as the PSNR. To test this statement, we computed maps of the absolute error between the averaged and denoised slices and the noise-free slices, as well as the mean absolute error (MAE) values. Figure 18 shows the case of SNR = 3 motionless images. See Appendix Figure A30 for other SNR values in the motionless case.

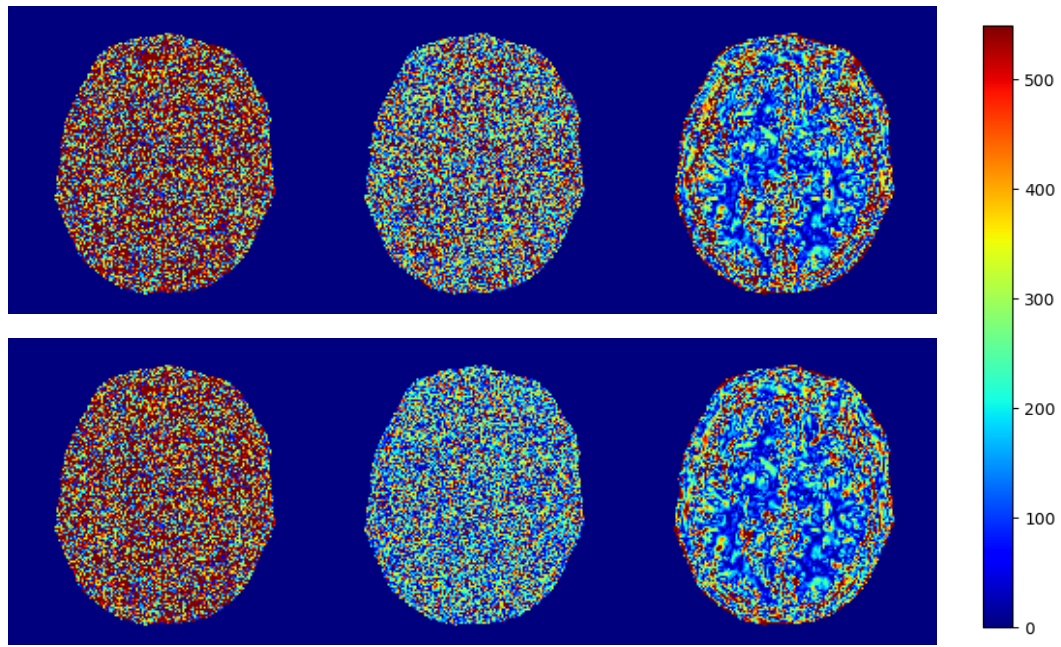

**Figure 18. Upper row: the model without the PSF, from left to right: NEX=1 noisy image (MAE=250.2), NEX=3 averaged image (MAE=147.75), CNN-denoised image (MAE=124.71). Lower row: the model with the PSF, from left to right: NEX=1 noisy image (MAE=250.15), NEX=5 averaged image (MAE=114.97), CNN-denoised image (MAE=105.61).**

It can be seen that for noisy averaged images, the error is distributed like the noise, but is lower as NEX increases. In the case of the denoised image, the error (within the tissue) is concentrated in the areas of edges, which seem to be more difficult to denoise. When the PSF is applied, the edges are smoothed and the denoised image is closer to the reference, which is also reflected by the increase of the PSNR-equivalent NEX from 3 to 5. However, the error in the denoised images is still concentrated at edges, i.e. the high-frequency image components. These observations are also valid for other SNR levels.

Another test that we performed on the denoised images was based on the assumption, that noise is randomly distributed with zero mean. If so, it can be reduced by averaging. Therefore, we prepared 100 instances of the same slice, each time affected by noise of the same distribution, but a random sample. In this case, due to the large number of repeated instances, we added pseudorandom noise with Gaussian distribution, zero mean and unit standard deviation to complex images. Then, we performed denoising on each of the 100 noisy instances. For each, we recorded not only the denoising result, but also the noise estimate. These estimates were then averaged. This way, we separated the zero-mean noise denoised by the CNN from deterministic image content that was filtered out, as shown in Figures 19 and 20.

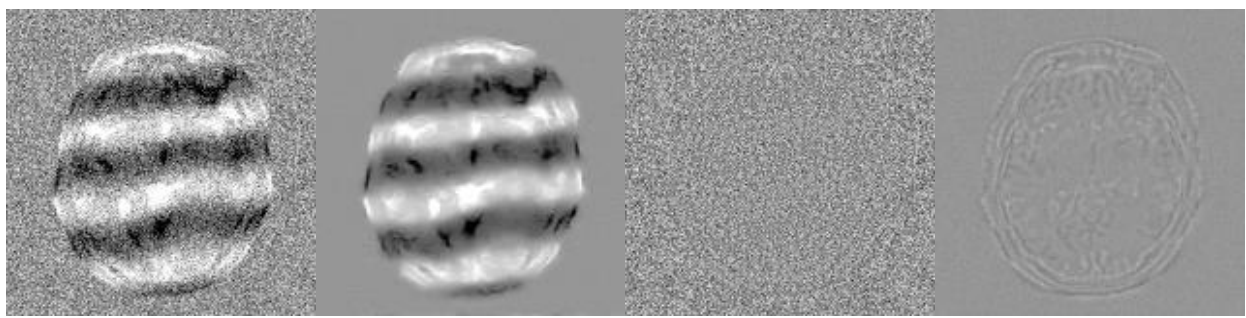

**Figure 19. From left to right: noisy image (NEX=1), CNN-denoised image, estimated noise map, average of 100 estimated noise maps. Simulated image, real part, SNR = 3. See Appendix Figure A31 for other SNR values.**

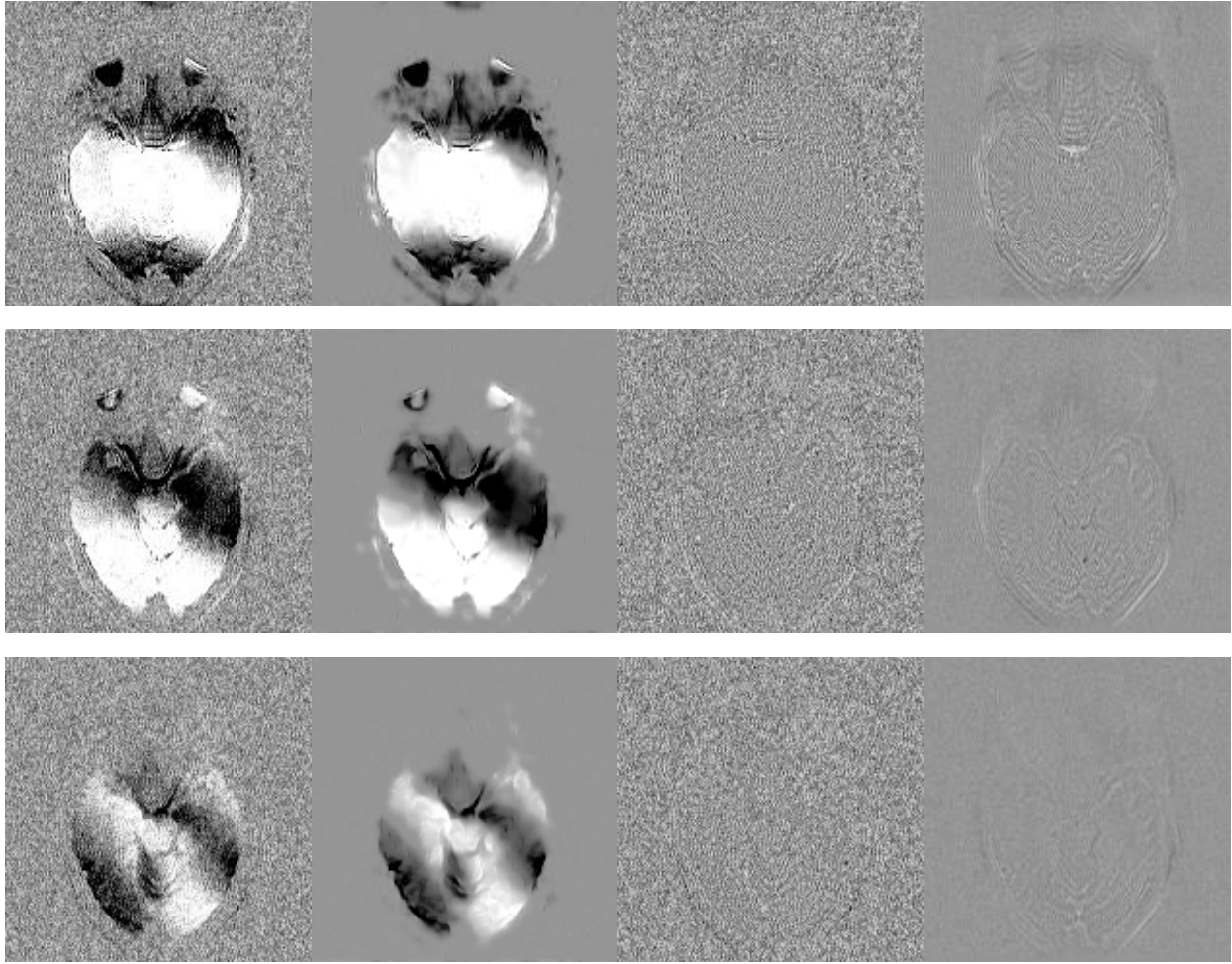

**Figure 20. From left to right: noisy image (NEX=1), CNN-denoised image, estimated noise map, average of 32 estimated noise maps, co-registered before averaging. From top to bottom: head DWI, b=0, 300x, 1000x, coil 1.**

The average of estimated noise maps (100 in the case of the simulated image, 32 for the real DWI image) reveals image edges erroneously subtracted from the noisy image. Noticeably, the intensity of the removed edges is mainly below the noise level, but they are able to be recognized by visual inspection of the individual estimated noise maps. In the real DWI (Figure 20), the removed edges have similar intensity as in simulated images. With increasing b-value, however, the intensity of removed edges decreases, presumably due to a lower high-frequency image content. Some of the edges are artificial edges created via Gibbs ringing. Removing them is of interest in DWI processing [Tax2022].

Since measuring the standard deviation of noise in the images is troublesome due to varying underlying image intensity, we studied pixel intensity variations between the 100 noisy instances created using the simulated images. Since all instances are affected by identically distributed noise, the standard deviation of a pixel can be measured across the noisy instances. Using this property, we computed noise standard deviation maps, as shown in Figure 21.

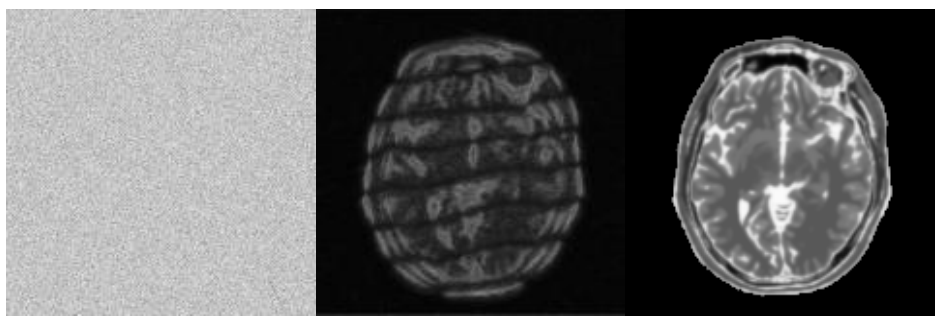

**Figure 21. Standard deviation maps of the real components of the noisy image with NEX=1 (left) and in the CNN-denoised image (middle), for the case of SNR=3. Wavy lines in the middle image reflect phase changes in the noisy image, which cause intensity drops. The right image is the modulus version of the same slice, shown for comparison. See Figure A32 in the Appendix for standard deviation maps of the denoised image at other SNR values.**

As visible in Figure 21, the standard deviation is uniformly distributed in the original noisy image, and then it is nonuniformly reduced, with the best result in the background, worse in the uniform regions of the tissue, and worst in at edges and in the regions of the close-to-maximum negative and positive intensity. We chose three pixels from a slices for this measurement – one located in the background, one in the mostly homogeneous area of the white matter and one in the high-intensity edge. We plotted their intensity across the 100 instances, as shown in Figure 22.

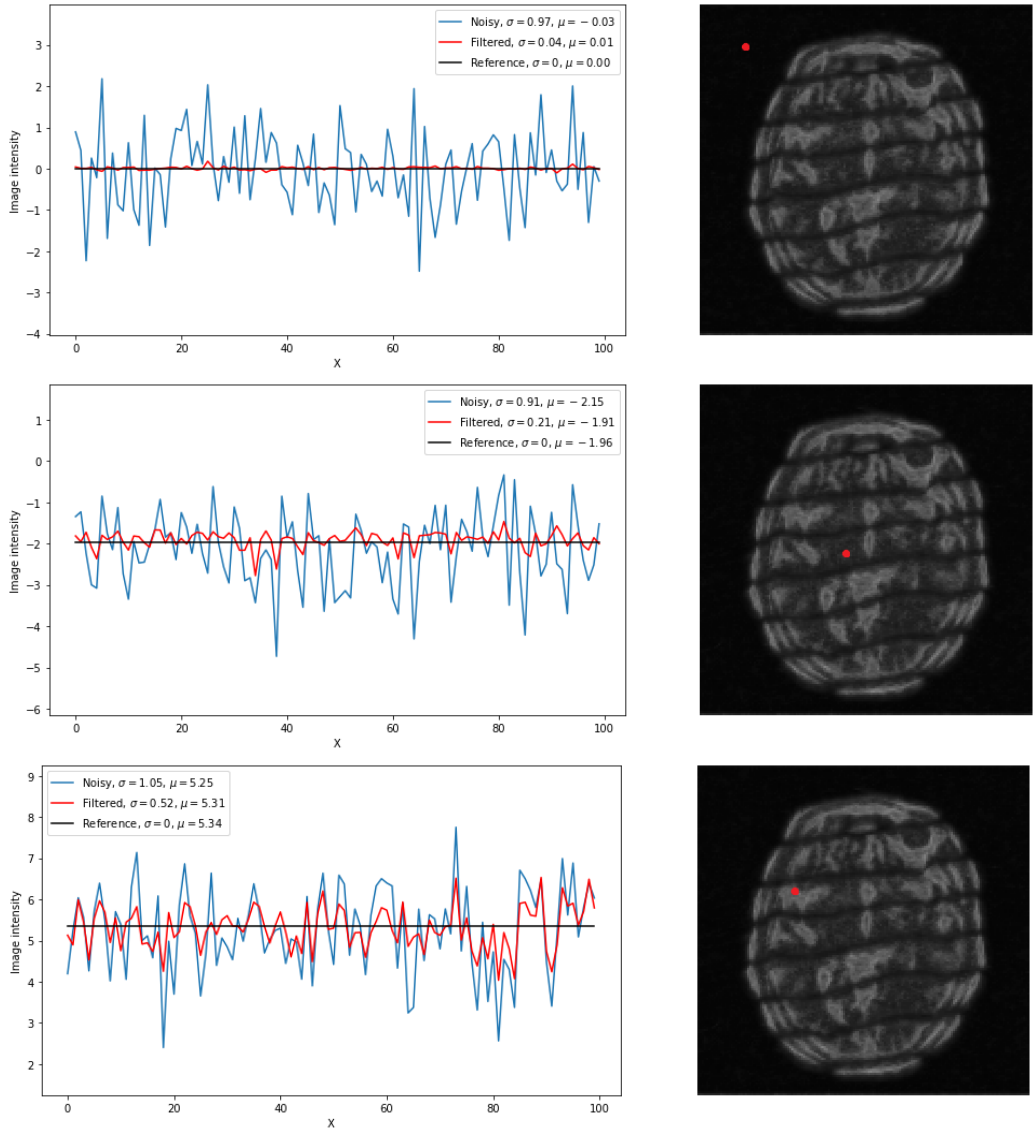

Figure 22. Left: Intensity plots for three chosen voxels: one located in the background, one in the mostly homogeneous area of the white matter and one in the high-intensity edge, across 100 noisy instances, SNR = 3. Noisy: blue line, denoised: red line. Right: maps of standard deviation. Red dots mark voxels for which the intensity plots were drawn. See Figures A33-A36 in the Appendix for other SNR values.

Analysing Figure 22, we can say that the noise was almost completely removed in the background. For the second voxel (white matter), the standard deviation was reduced to 0.21 (from 0.91, due to a small noise sample not equal exactly 1) and also the mean was closer to the true voxel intensity (-1.91 vs. -2.15 with -1.96 being the true value). The voxel located at the edge showed the standard deviation of 0.52, but its mean was also closer to the reference then the mean from the noisy image (denoised 5.31 vs. noisy 5.25 with 5.34 being the true value). The ratios of standard deviations for these three voxels are 24.25, 4.33 and 2.02, respectively. Such gain in the SNR would require NEX values of 588, 18 and 4.

We conclude the analysis of the denoising-introduced distortions by studying the denoised trace DWI images of the head. As mentioned in the Section 4.2., the ADC map calculated from denoised DWI shows improperly high ADC values near the skull and in the eyeballs. We draw intensity profiles in a horizontal line of a chosen slice, crossing the skull, gray matter, white matter and the cerebellum, in the b0, b300 trace and b1000 trace images (Figure 23).

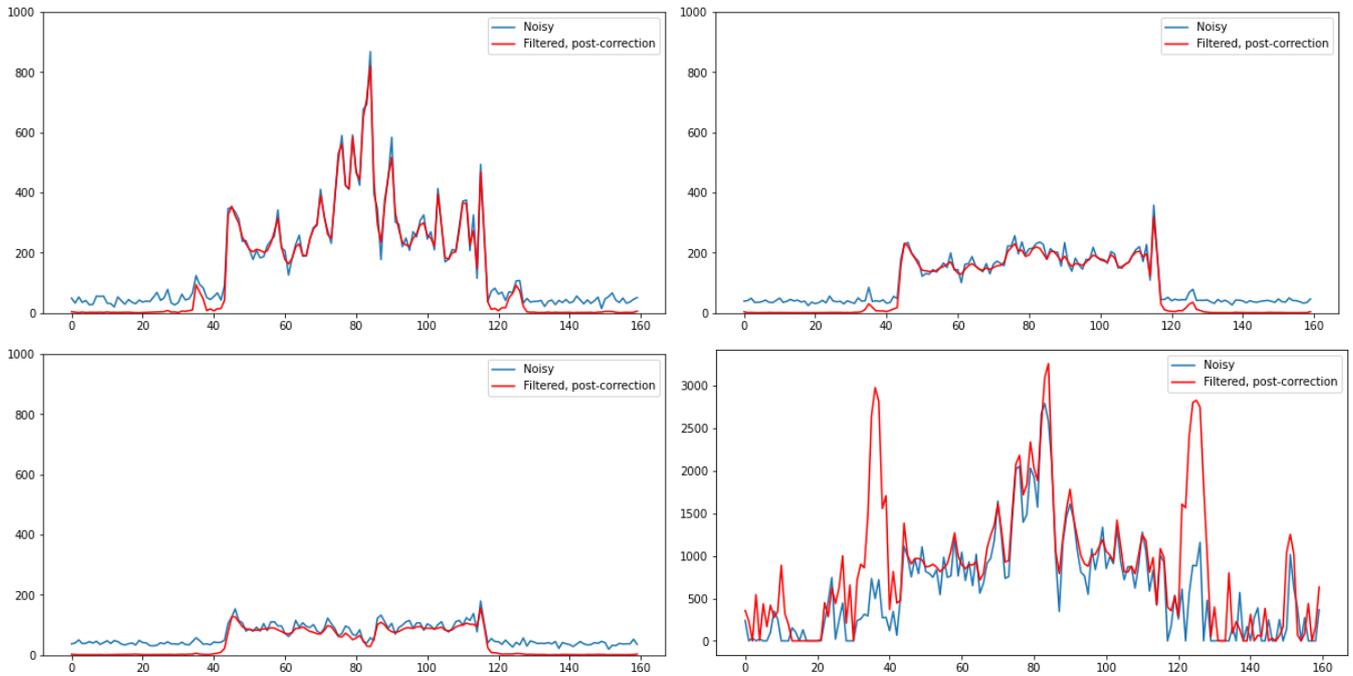

Figure 23. Intensity profiles drawn at the 120[th] of 160 rows of the slice shown in Figures 15, 16 and 17 – for noisy NEX=1 and denoised b=0, b=300 trace, b=1000 trace DWI and the corresponding ADC map.

The profiles reveal the source of the observed artifacts. In the skull region (around 35 and 125 on the horizontal axis of the profiles), one can notice significant filtration. The skull is filtered out completely in the b=1000 images and overly in b=300 images, which lead to an artifactual increase of the ADC in this region. The same happens in the region of the eyeball, which are filtered out overly in b=1000 images, as seen in Figure 16. Because the brain tissue has higher intensity than the skull or eyeball in high-b-value images, we do not observe this artifact in this crucial region, while observing significant reduction of noise as compared to NEX=1, 9 and 25 averaged images.

## 5. Discussion

Even though denoising is a difficult problem, evaluating the quality of images is demanding as well. In our experiment, similarly to many other works related to image enhancement, we used well known, comparative metrics: SSIM and PSNR. Using these metrics, we estimated the NEX necessary to obtain, through averaging, similar quality as by denoising a single image. However, two images with the same or close SSIM or PSNR may be different images in details, because these metrics are global. As we observed, denoised images had much less noisy appearance than the images with similar metric values. As the collected evidence suggests, this is due to the errors introduced by the CNN filter, or, using a different terminology, the bias of the true intensity estimator. This is why in addition to studying the quality metric values we analysed the denoising results by denoising 100 identical slices, with different samples of identically distributed noise. This showed that the CNN filter indeed significantly reduced the standard deviation, but the estimated voxel intensity was limited by the bias of the filter, and varied depending on the voxel neighbourhood. In turn, averaging images in the complex domain, although less efficient than CNN denoising, is free of bias and uniformly reduces the noise standard deviation, provided that they are motionless. This is why at some NEX level the denoising results were outperformed by averaging. However, when motion was taken into account in the simulation process, denoising was not outperformed by averaging in most cases, because the blurring and artifacts introduced by averaging misplaced images were greater than the CNN filter bias. We observed that averaging motion-corrupted images had less significant effect on image PSNR than noise for the few first NEX values (Figure 11), but for higher values the quality dropped. Thus, the CNN filter allowed to obtain metric values non achievable by averaging, which was shown for SNR = 5, 7 and 9 for the SSIM and SNR = 3, 5, 7 and 9 for the PSNR. With this stated, it should be noticed that even if a certain NEX number outperforms denoising, it is possible to denoise several repeated images and average them afterwards. Although we did not evaluate such a strategy in this paper, we tried this out to see increased quality when averaging denoised images.

The discrepancy between PSNR and SSIM is another factor that complicated evaluation. We computed both metrics, but relied more on the PSNR, which allows straightforward analysis of local errors. With this metric, we observed that at least NEX=2 was necessary to match denoising quality for the case of high-resolution images without motion or PSF. When PSF was accounted for, this NEX value increased to 3. Although these values may seem low, the difference between NEX=2 and NEX=1 imaging is twice shorter imaging time, and thrice shorter for NEX=3.

We also noticed that, for greater SNR values and motionless repeated images, the efficiency of using the CNN filter decreased compared to averaging. For example, it was enough to average 3 repeated images with SNR = 9, while NEX = 8 was necessary with SNR = 3 (according to the PSNR). One conclusion could be that the CNN reduced noise standard deviation to a similar extent independent of the SNR, and the NEX required to decrease noise standard deviation by averaging drops with growing SNR. On the other hand, we noticed that the image details erroneously filtered out had higher intensity when SNR was larger. As such, the CNN filter might induce more distortion compared to the noise for higher SNR levels. Finally, this effect might be attributed to an obvious feature of ARI. The SNR gain is known to be the square root of the NEX. For SNR = 3 and NEX = 2, the gain is approximately 1.41 which improves the SNR to 4.23. For SNR = 9 and the same value of NEX, the gain is also 1.41, but improves the SNR to 12.69. The neural network seems not to follow such a property. Similar network properties were noticed in [Jurek2020], where the CNN was more efficient in the resolution improvement task, compared to averaging of interpolated images, for larger slice thicknesses.

In our experiments, we compared, both for simulated and real images, denoising results performed before and after corrections necessary for EPI data. Quantitative and visual evaluation showed superiority of the latter option. Denoising the image before correction of non-uniform sampling and Nyquist ghosts results in error after ghost correction. A low-intensity noise was visible at the location of the ghosts in this case.

For denoising real DWI, we used three networks, trained on data with different SNR, dependent on the b-value. We tried to adapt the network target SNR by computing the SNR coil-wise for each b-value and even slice-wise, since the mean slices SNR values might be different. However, denoising results were not noticeably different, so we chose the simpler, non-adaptive option.

As it was mentioned before, we chose a simple CNN architecture and a relatively small training dataset. This allows fast training (times around 15 minutes) and still leads to significantly denoised images. Probably, better results can be obtained with a larger training dataset (BrainWeb offers more than one digital brain maps, data augmentation can be used as well) and a more sophisticated architecture, if necessary. The main problem identified was the removal of noise together with low-intensity, but non-negligible edge content. Logical analysis of the training process leads to a conclusion, that this is caused by the choice of the type of the loss function, specifically the mean squared error. This function is blind to everything except the difference between the intensities, i.e. it cannot distinguish edges from noise. This way large noise reduction was achieved at the cost of removing some image content of intensity below the noise level, even if the SNR was large. Finding the solution to this problem remains in our main scope of interest and we will continue the research in this direction.

Improving noise filtration results can also be obtained in other ways. Fadnavis et al. took into account that multiple-b-value acquisitions are related to each other and some constraints may be necessary during denoising, so that the results remain correct for DWI modelling [Fadnavis2020]. In our work, we neglected these relation and we performed denoising independently for each complex component of each image and receiver coil. The ADC maps we obtained are affected by artifacts in the region of the skull and eyeballs. Although taking into account the relations between the acquired multiple b-value images may be relevant, we rather see the reasons for the observed artifacts in the very low SNR level in the skull and eyeballs in b=1000 images of individual coils. Because filtration is performed before combining the complex components and coil images, low intensity tissue signals may be removed or significantly attenuated together with noise. Alternatively, filtration could be performed in the modulus domain in multicoil images, were the SNR is increased, but the noise distribution may be more complicated. We will test this possibility in future works, however, it is not guaranteed that low intensity structures will be visible in NEX=1 images. This means that in some cases it might be necessary to perform imaging with a small NEX number greater than 1. This would be associated with a relatively small risk of patient movement, but would increase the SNR. After all, we observed the artifacts only outside the brain, which is the actual organ of interest in this imaging domain. In the close future, we plan to use a diffusion phantom to quantitatively assess the ADC values obtained from denoised images.

Our EPI data model included typical EPI artifacts such as ghosting and non-uniform sampling effects, as well as phase variations and noise. All of these effects were applied to the simulated spin-echo T2-weighted brain image and were data-driven, being derived from the real DWI images and air scans. We neglected other common artifacts such as the geometric distortion, since they do not affect the noise or the denoising procedure. In the future, however, we might extend the data model to better fit a multi b-value EPI DWI dataset. It is possible to upgrade the BrainWeb phantom by defining the ADC value of each tissue in the phantom, which will allow to obtain images with b>0. It is also possible to derive receiver coil profiles from the acquired DWI data to simulate multicoil imaging. Improving the data model would allow to draw more conclusions about the denoising performance and point out the optimal stage in the DWI reconstruction and processing pipeline at which filtration is most beneficial.

The presented method is not limited to brain, EPI or DWI data, although our simulation was focused so. A similar approach can be used to denoise other low SNR MRI images, for example functional brain MRI or dynamic contrast-enhanced MRI of kidneys. Reducing the necessary NEX may either shorten examination times or allow for higher resolution within the same study time and equivalent image quality. We plan to apply the method to denoise functional brain MRI images, abdomen DWI and pelvis DWI and evaluate the quality of denoising of real images in a real-world image interpretation problem, such as functional brain mapping or radiological interpretation.

## 6. Conclusions

We demonstrated and analysed an image denoising method based on a convolutional neural network and transfer learning and targeted at DWI images. We showed that denoised images are significantly closer to a clean reference than averaged images. This method can be used to obtain good quality DWI in a shortened examination time, but further research is necessary to improve denoising results and investigate the clinical utility of the denoised real MRI images and ADC maps.

# Appendix to Supervised Denoising of Diffusion-Weighted Magnetic Resonance Images Using a Convolutional Neural Network and Transfer Learning

Jakub Jurek, Andrzej Materka || Kamil Ludwisiak, Agata Majos || Kamil Gorczewski, Kamil Cepuch, Agata Zawadzka

Additional PSNR and SSIM plots for Section 4.1.1.

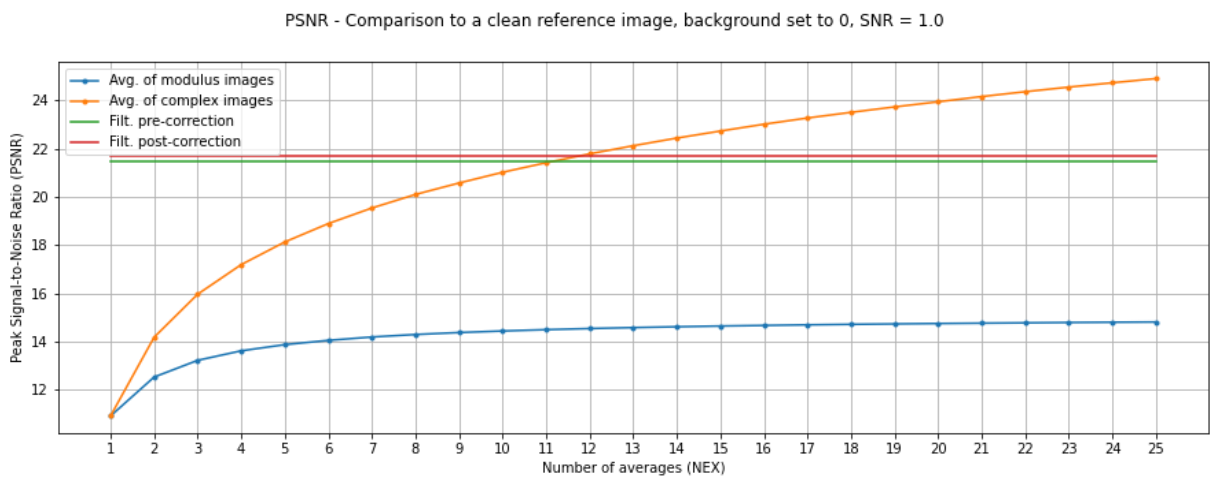

**Figure A1. PSNR plot versus the NEX for the averaged images with SNR = 1. Pre- and post-correction denoised NEX = 1 image PSNR vales (for tissue) are shown as horizontal lines. Model with PSF and no motion.**

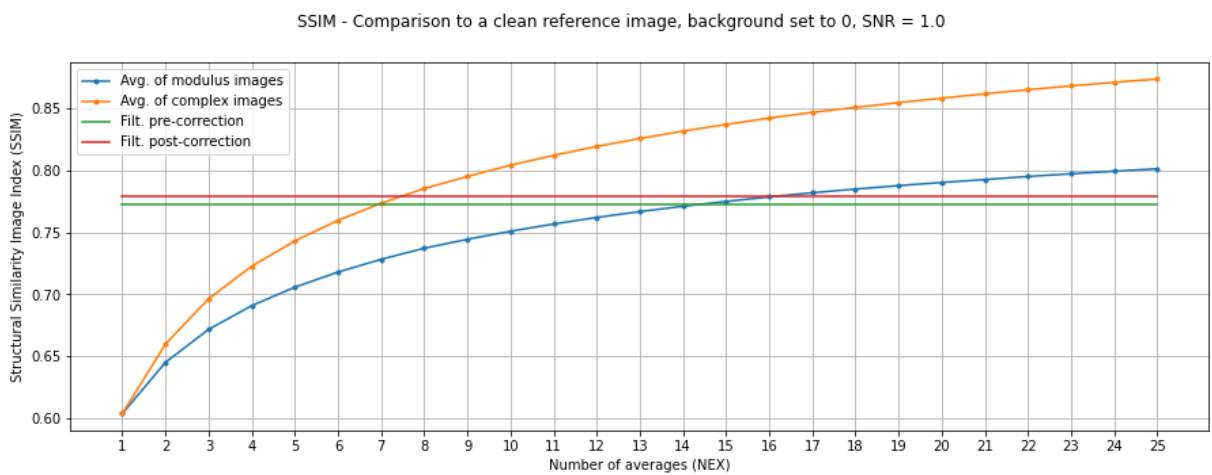

**Figure A2. SSIM plot versus the NEX for the averaged images with SNR = 1. Pre- and post-correction denoised NEX = 1 image SSIM vales (for tissue) are shown as horizontal lines. Model with PSF and no motion.**

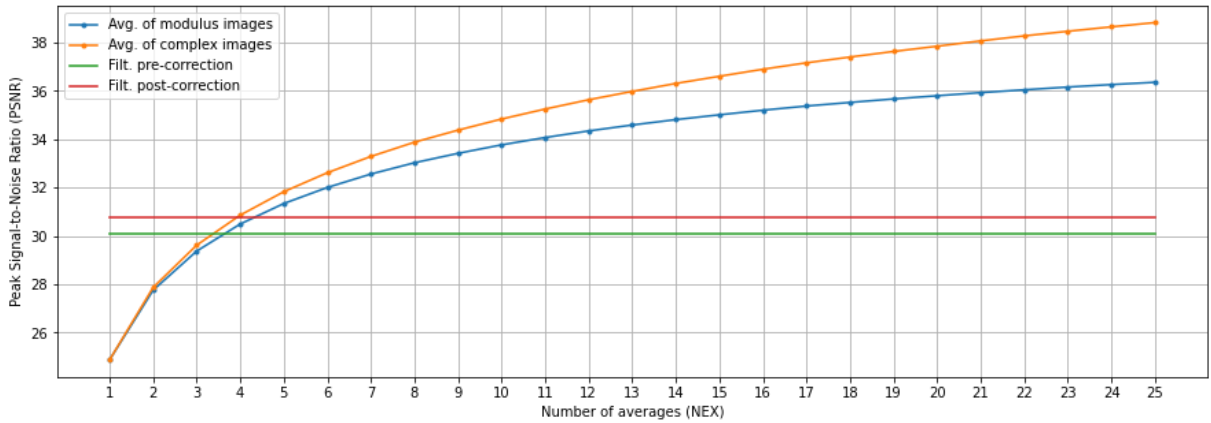

**Figure A3.** PSNR plot versus the NEX for the averaged images with SNR = 5. Pre- and post-correction denoised NEX = 1 image PSNR vales (for tissue) are shown as horizontal lines. Model with PSF and no motion.

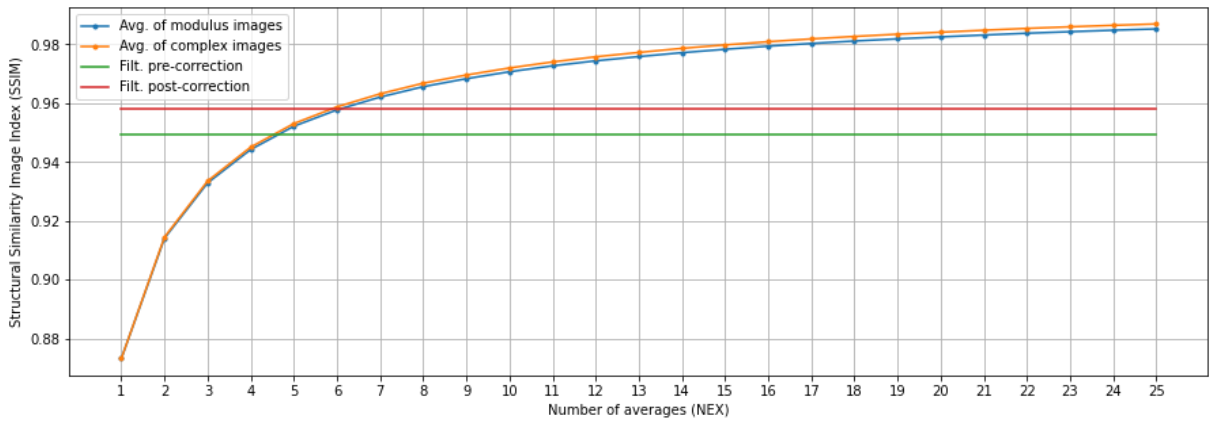

**Figure A4.** SSIM plot versus the NEX for the averaged images with SNR = 5. Pre- and post-correction denoised NEX = 1 image SSIM vales (for tissue) are shown as horizontal lines. Model with PSF and no motion.

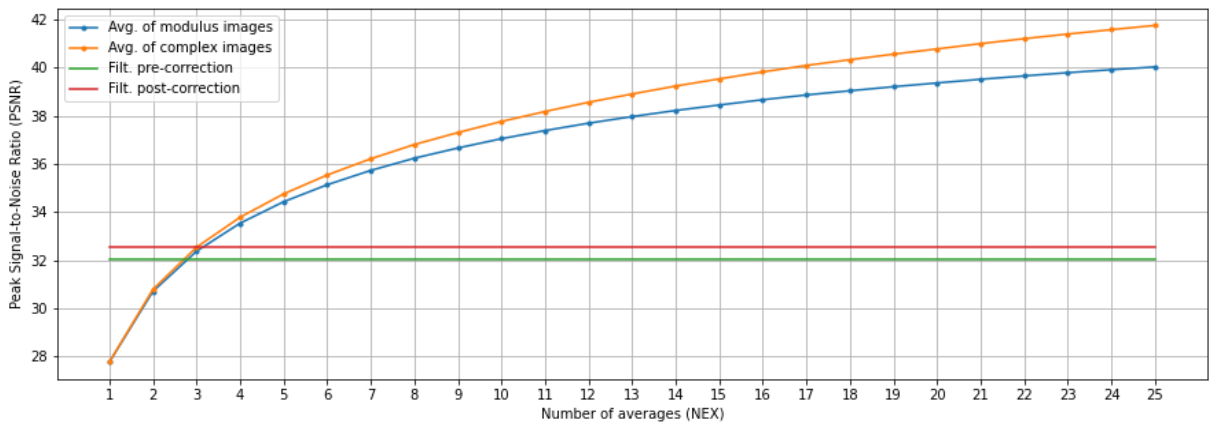

**Figure A5. PSNR plot versus the NEX for the averaged images with SNR = 7. Pre- and post-correction denoised NEX = 1 image PSNR vales (for tissue) are shown as horizontal lines. Model with PSF and no motion.**

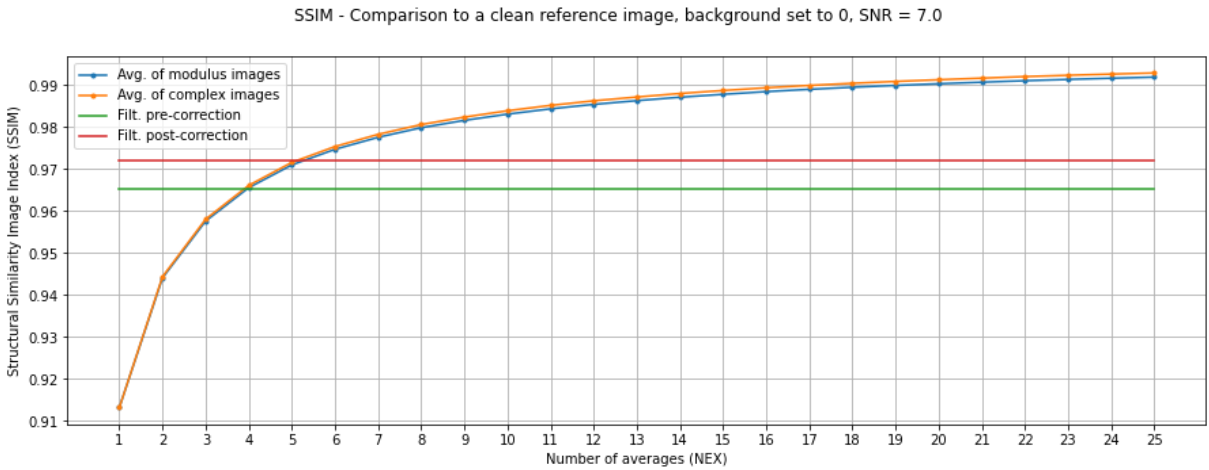

**Figure A6. SSIM plot versus the NEX for the averaged images with SNR = 7. Pre- and post-correction denoised NEX = 1 image SSIM vales (for tissue) are shown as horizontal lines. Model with PSF and no motion.**

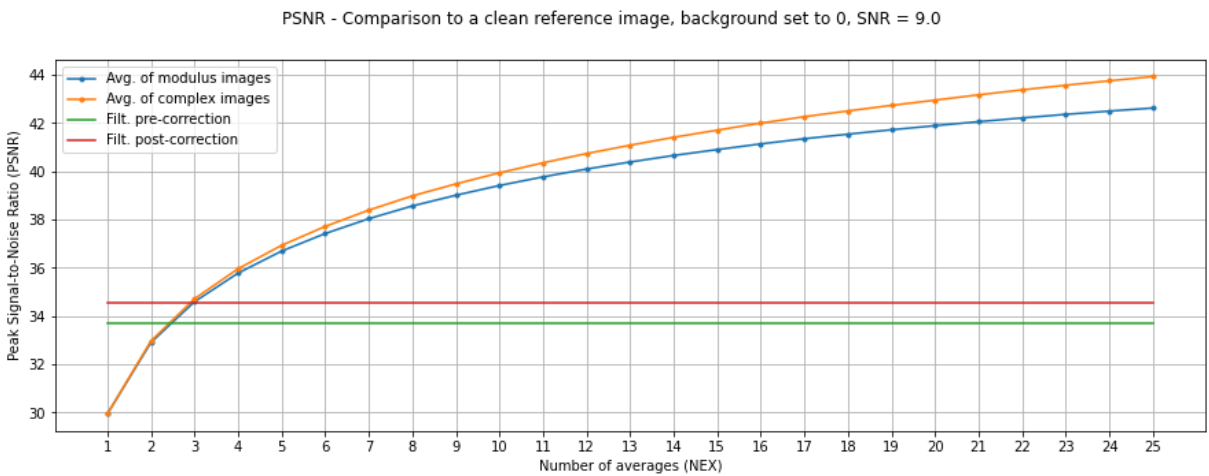

**Figure A7. PSNR plot versus the NEX for the averaged images with SNR = 9. Pre- and post-correction denoised NEX = 1 image PSNR vales (for tissue) are shown as horizontal lines. Model with PSF and no motion.**

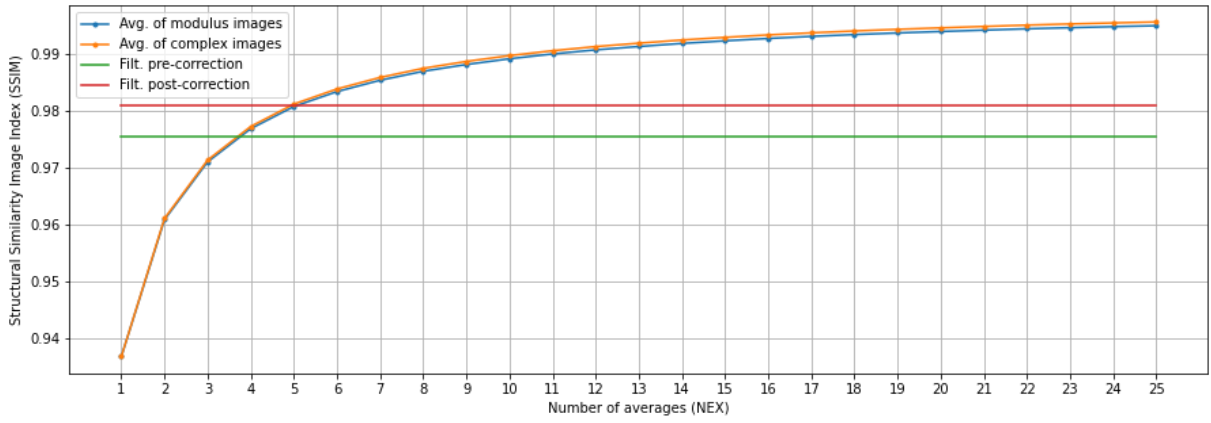

**Figure A8.** SSIM plot versus the NEX for the averaged images with SNR = 9. Pre- and post-correction denoised NEX = 1 image SSIM vales (for tissue) are shown as horizontal lines. Model with PSF and no motion.

Averaged images with varying SNR and NEX value – to Section 4.1.1.

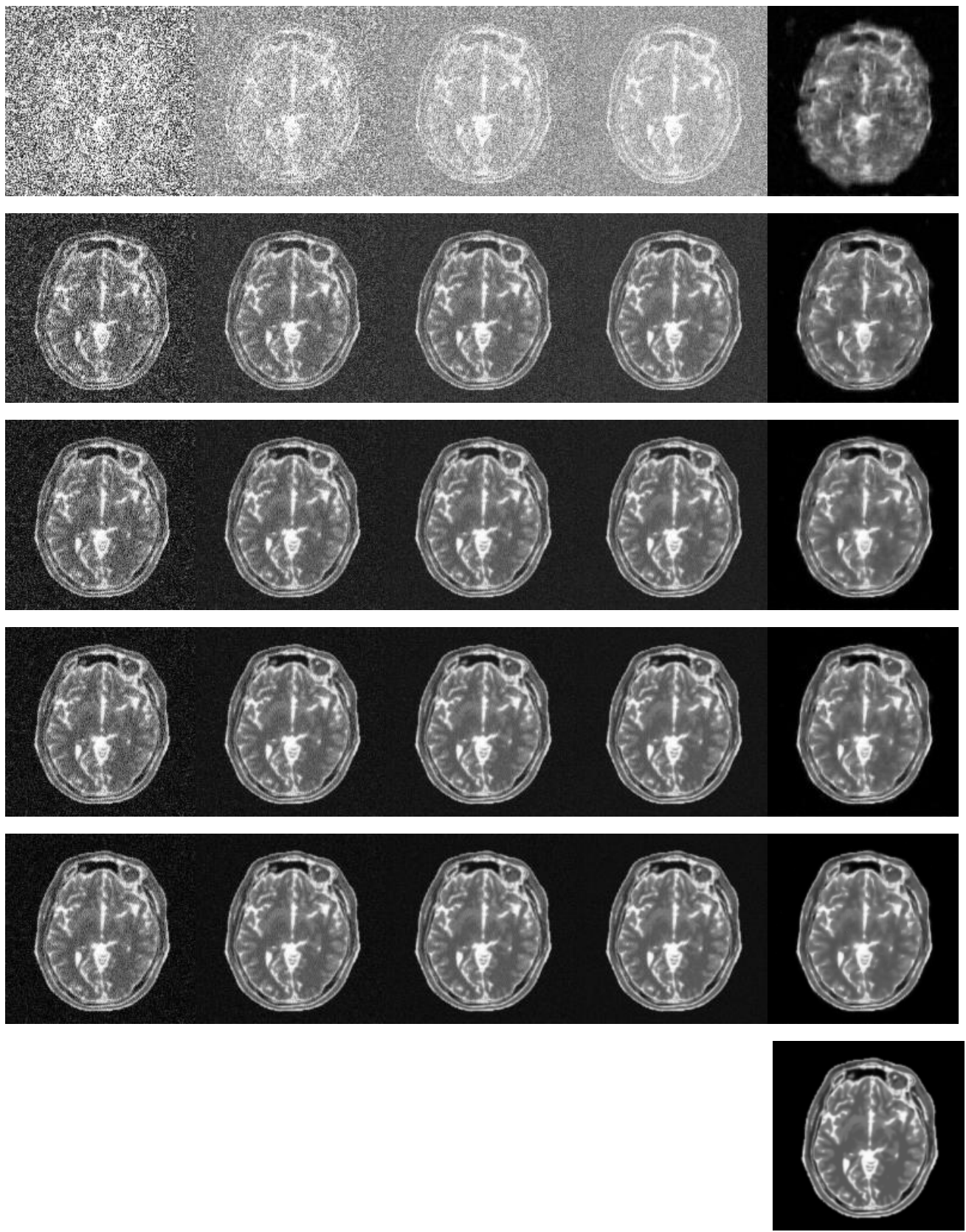

**Figure A9.** From left: Repeated images (ARI) averaged in the modulus domain: NEX=1, 4, 9, 16, then a CNN-denoised image. From top to bottom: SNR=1, 3, 5, 7, 9, last row: clean image. Model with PSF and no motion

## Additional PSNR and SSIM plots for Section 4.1.2.

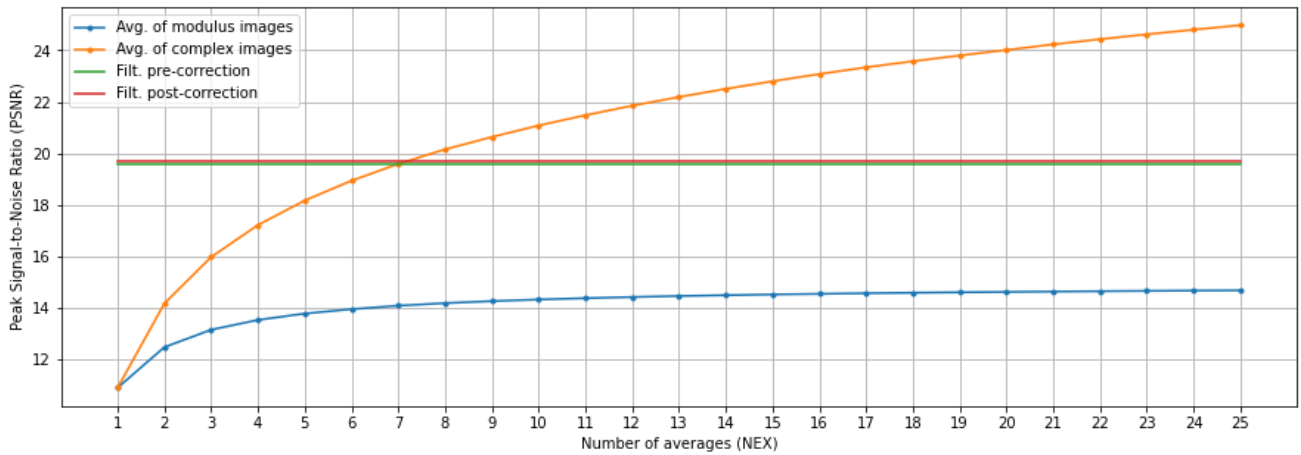

**Figure A10.** PSNR plot versus the NEX for the averaged images with SNR = 1. Pre- and post-correction denoised NEX = 1 image PSNR vales (for tissue) are shown as horizontal lines. Model with without PSF and motion.

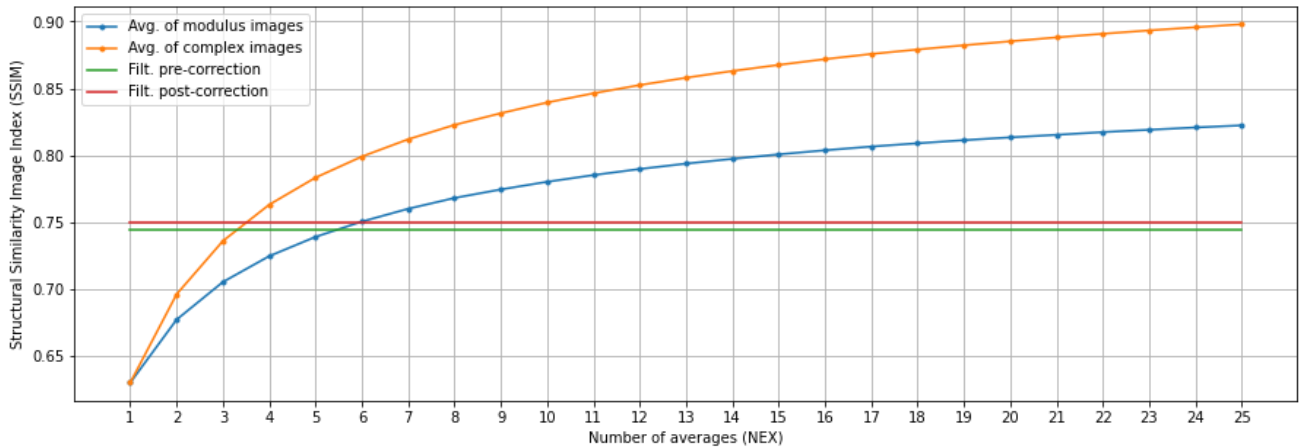

**Figure A11.** SSIM plot versus the NEX for the averaged images with SNR = 1. Pre- and post-correction denoised NEX = 1 image SSIM vales (for tissue) are shown as horizontal lines. Model with without PSF and motion.

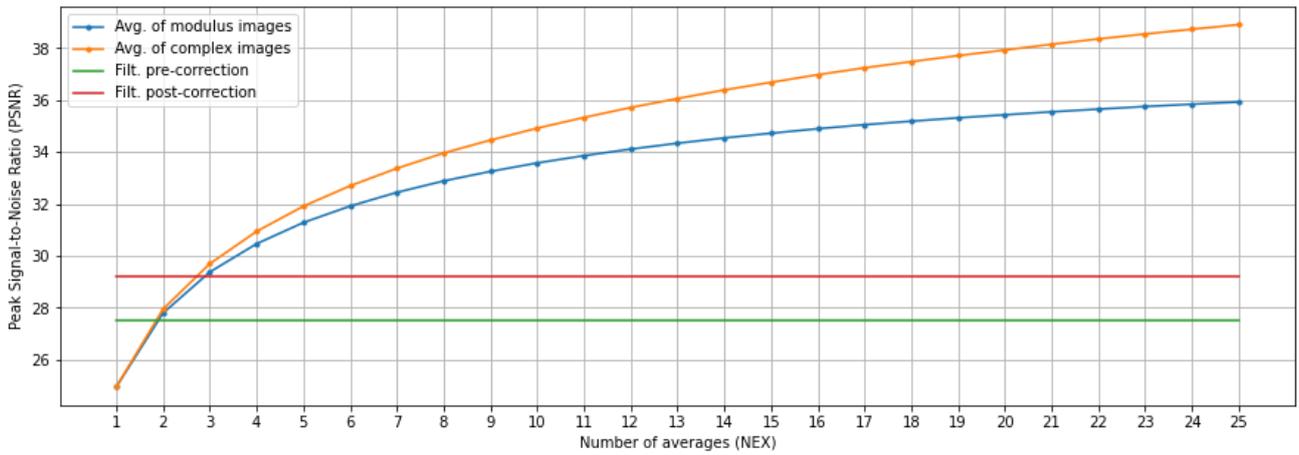

**Figure A12.** PSNR plot versus the NEX for the averaged images with SNR = 5. Pre- and post-correction denoised NEX = 1 image PSNR vales (for tissue) are shown as horizontal lines. Model with without PSF and motion.

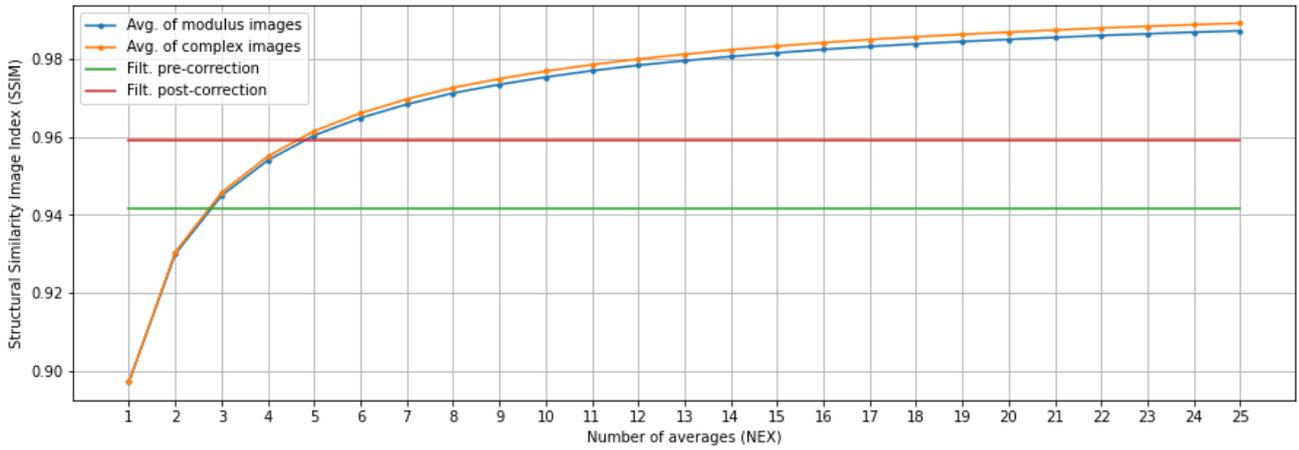

**Figure A13.** SSIM plot versus the NEX for the averaged images with SNR = 5. Pre- and post-correction denoised NEX = 1 image SSIM vales (for tissue) are shown as horizontal lines. Model with without PSF and motion.

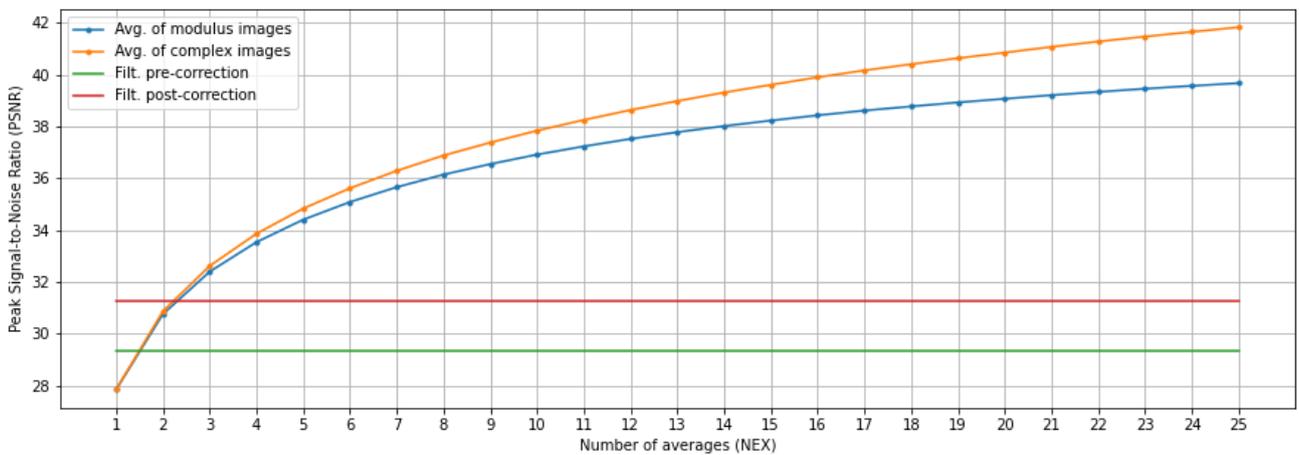

**Figure A14. PSNR plot versus the NEX for the averaged images with SNR = 7. Pre- and post-correction denoised NEX = 1 image PSNR vales (for tissue) are shown as horizontal lines. Model with without PSF and motion.**

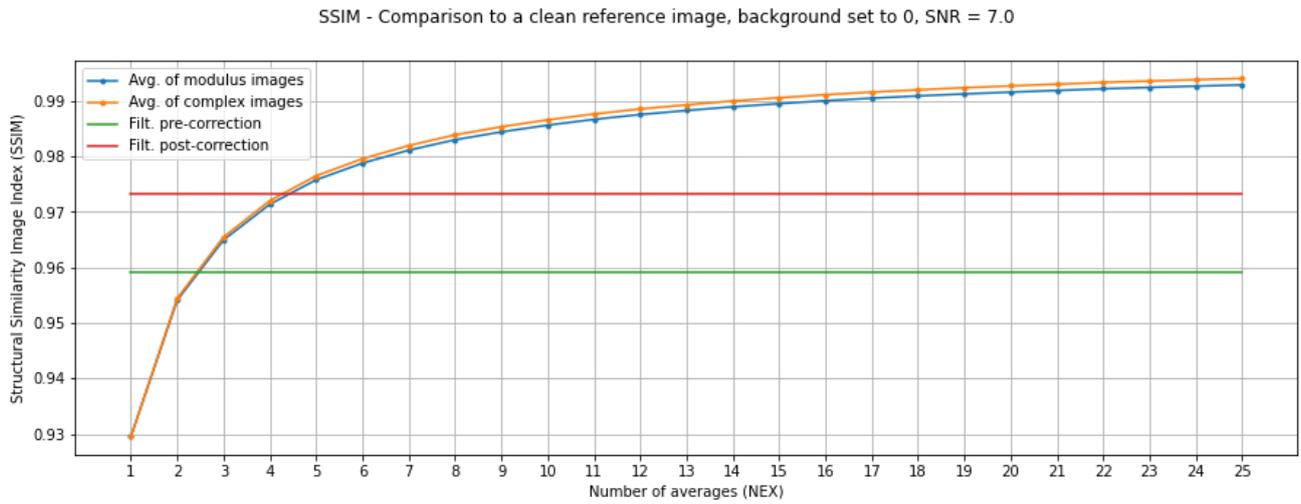

**Figure A15. SSIM plot versus the NEX for the averaged images with SNR = 7. Pre- and post-correction denoised NEX = 1 image SSIM vales (for tissue) are shown as horizontal lines. Model with without PSF and motion.**

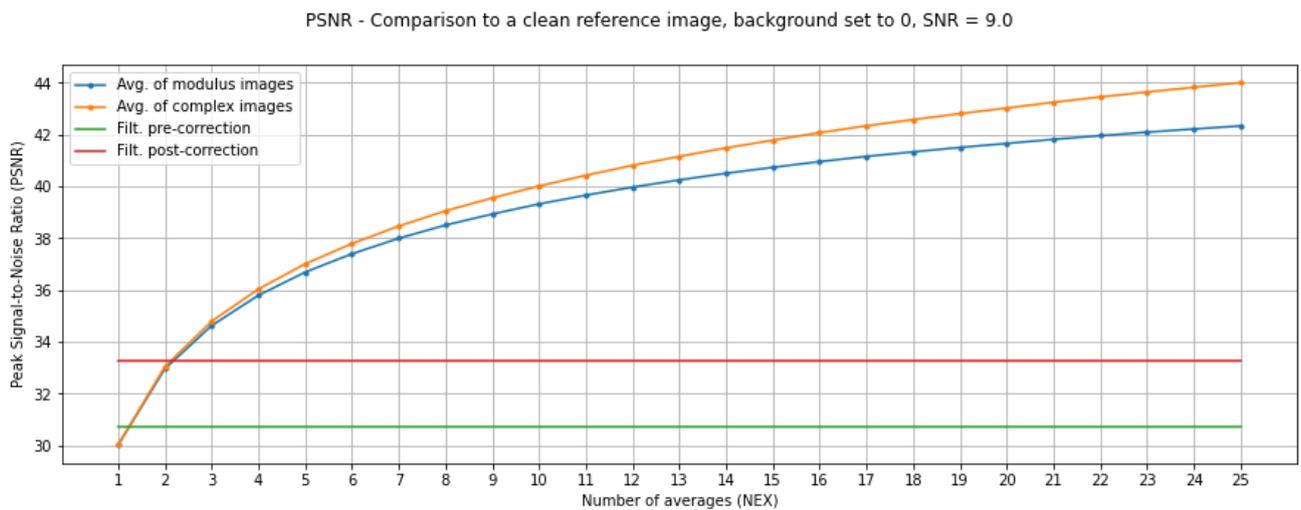

**Figure A16. PSNR plot versus the NEX for the averaged images with SNR = 9. Pre- and post-correction denoised NEX = 1 image PSNR vales (for tissue) are shown as horizontal lines. Model with without PSF and motion.**

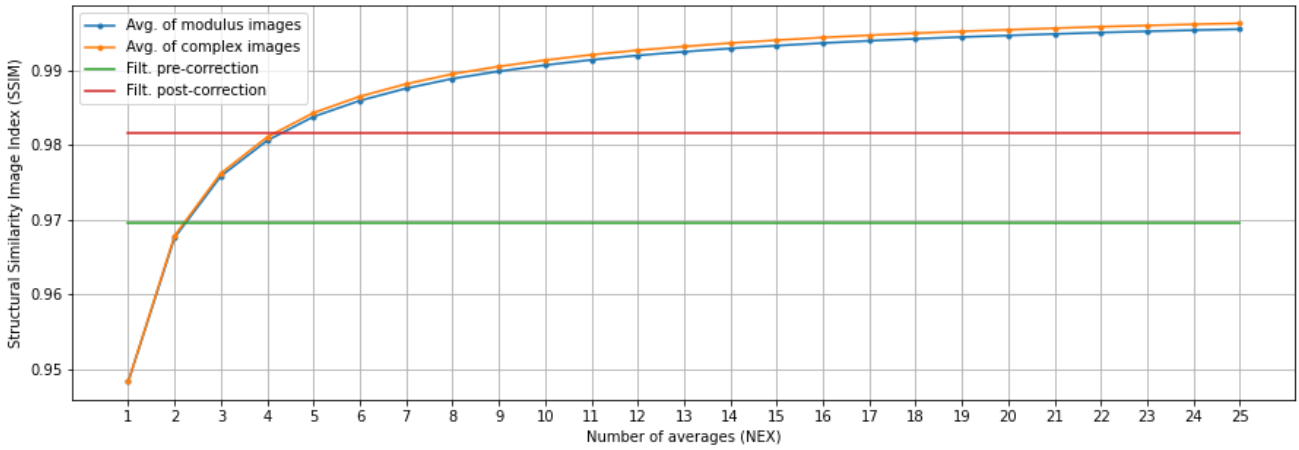

**Figure A17.** SSIM plot versus the NEX for the averaged images with SNR = 9. Pre- and post-correction denoised NEX = 1 image SSIM vales (for tissue) are shown as horizontal lines. Model with without PSF and motion.

## Additional PSNR and SSIM plots for Section 4.1.3.

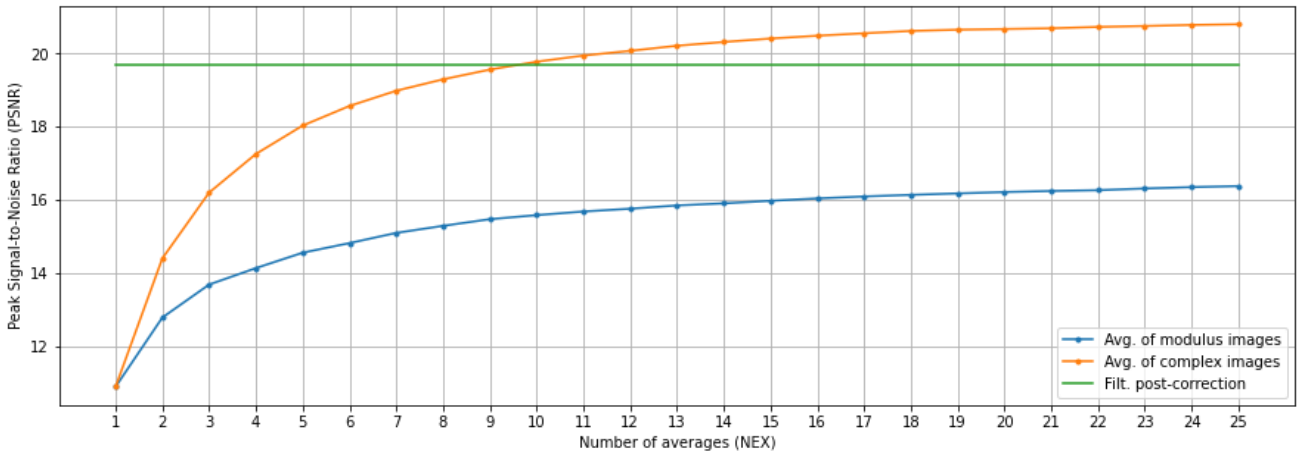

**Figure A18.** PSNR plot versus the NEX for the averaged images with SNR = 1. Pre- and post-correction denoised NEX = 1 image PSNR vales (for tissue) are shown as horizontal lines. Model with without PSF and motion.

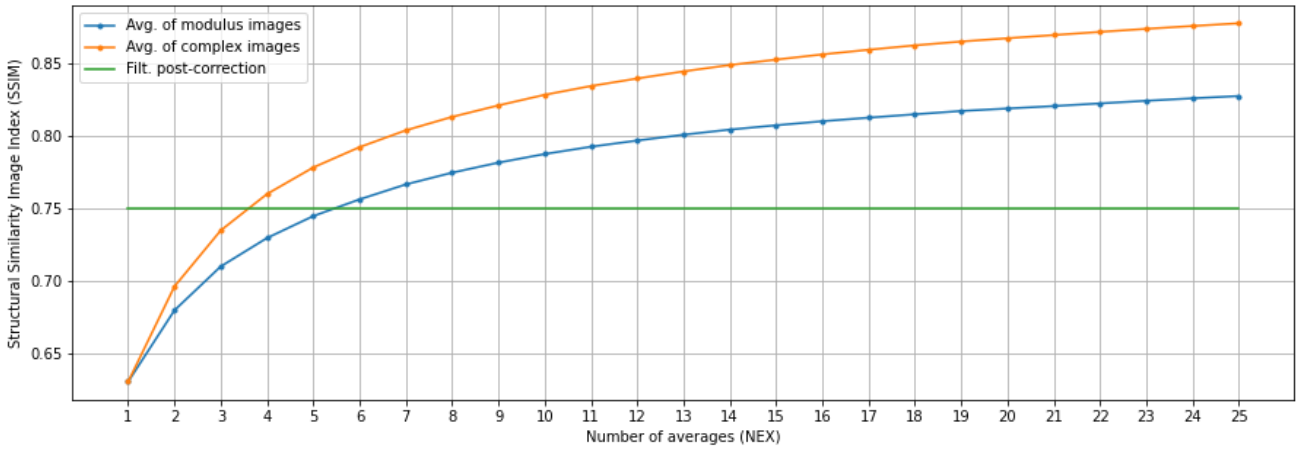

**Figure A19.** SSIM plot versus the NEX for the averaged images with SNR = 1. Pre- and post-correction denoised NEX = 1 image SSIM vales (for tissue) are shown as horizontal lines. Model with without PSF and motion.

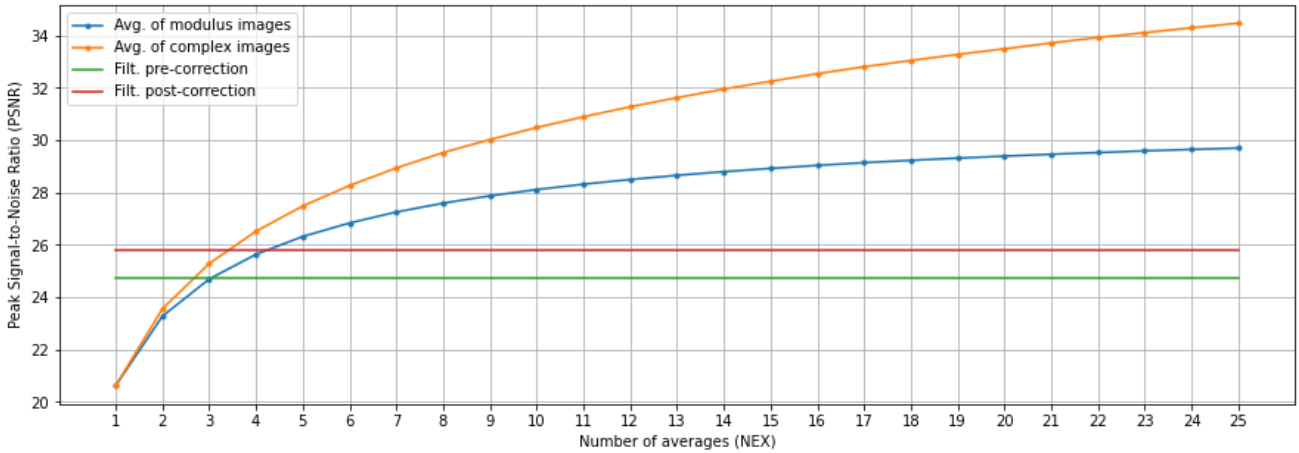

**Figure A20.** PSNR plot versus the NEX for the averaged images with SNR = 3. Pre- and post-correction denoised NEX = 1 image PSNR vales (for tissue) are shown as horizontal lines. Model with without PSF and motion.

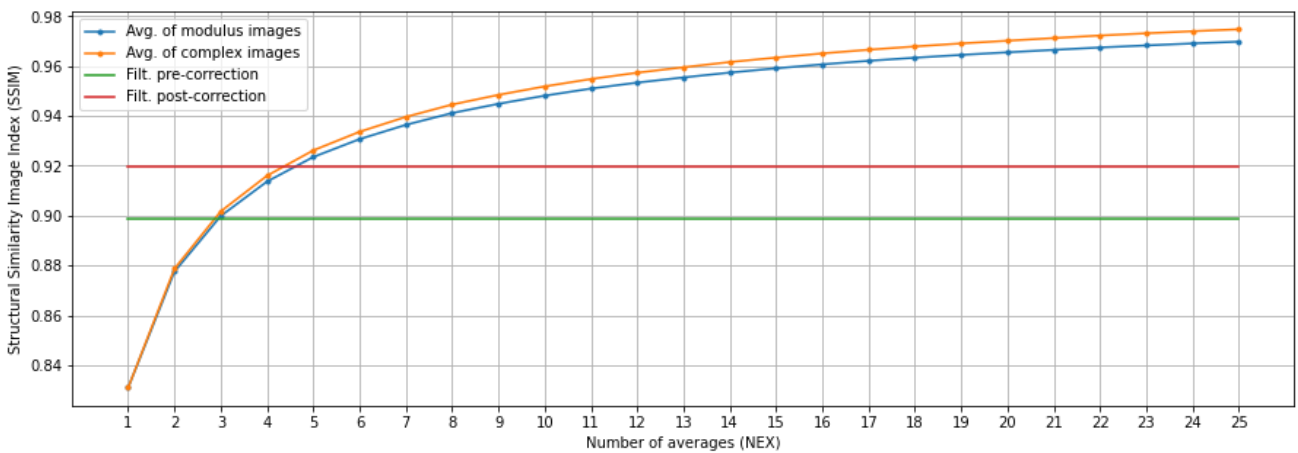

Figure A21. SSIM plot versus the NEX for the averaged images with SNR = 3. Pre- and post-correction denoised NEX = 1 image SSIM vales (for tissue) are shown as horizontal lines. Model with without PSF and motion.

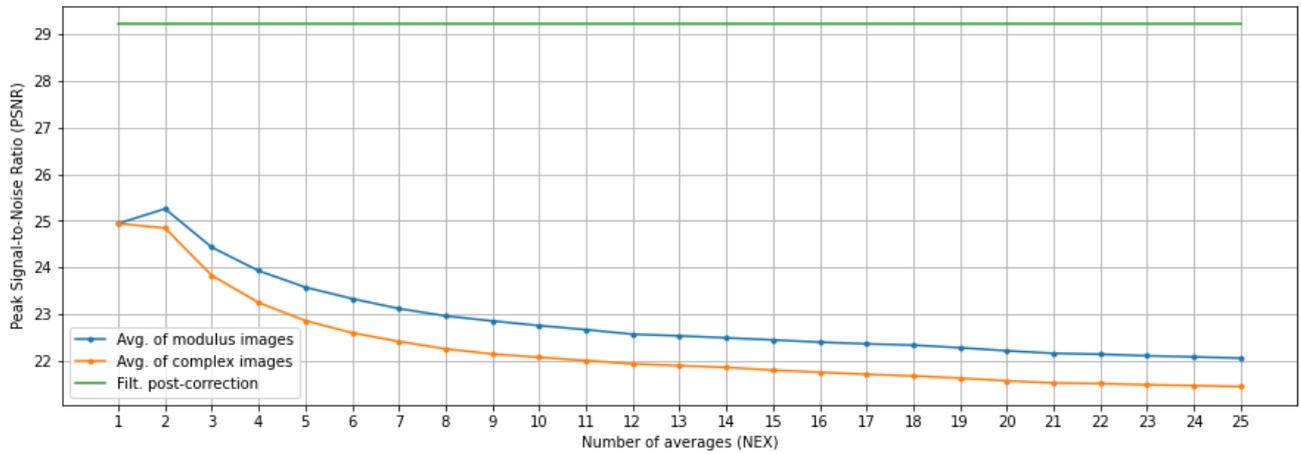

Figure A22. PSNR plot versus the NEX for the averaged images with SNR = 5. Pre- and post-correction denoised NEX = 1 image PSNR vales (for tissue) are shown as horizontal lines. Model with without PSF and motion.

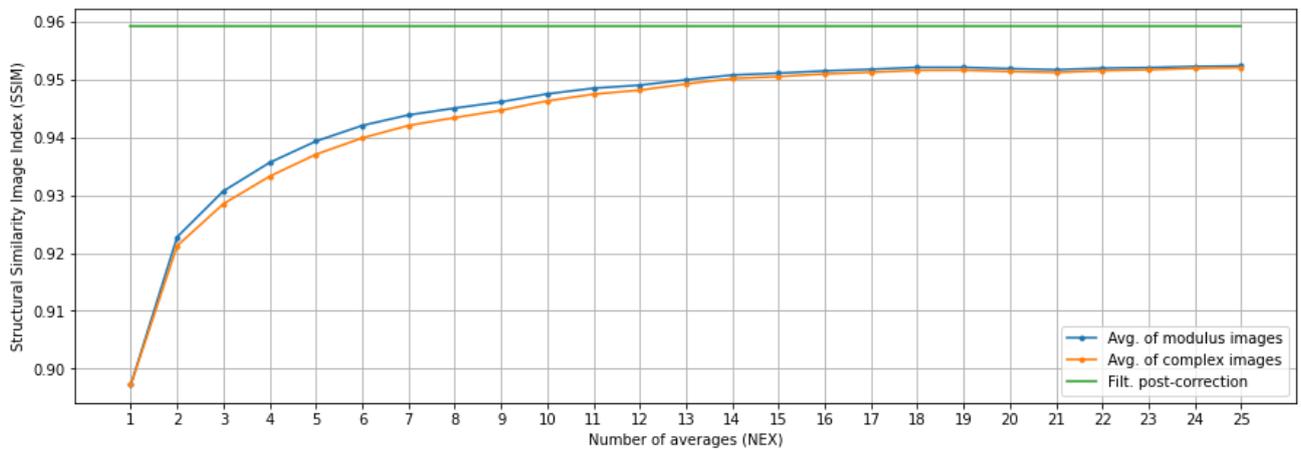

Figure A23. SSIM plot versus the NEX for the averaged images with SNR = 5. Pre- and post-correction denoised NEX = 1 image SSIM vales (for tissue) are shown as horizontal lines. Model with without PSF and motion.

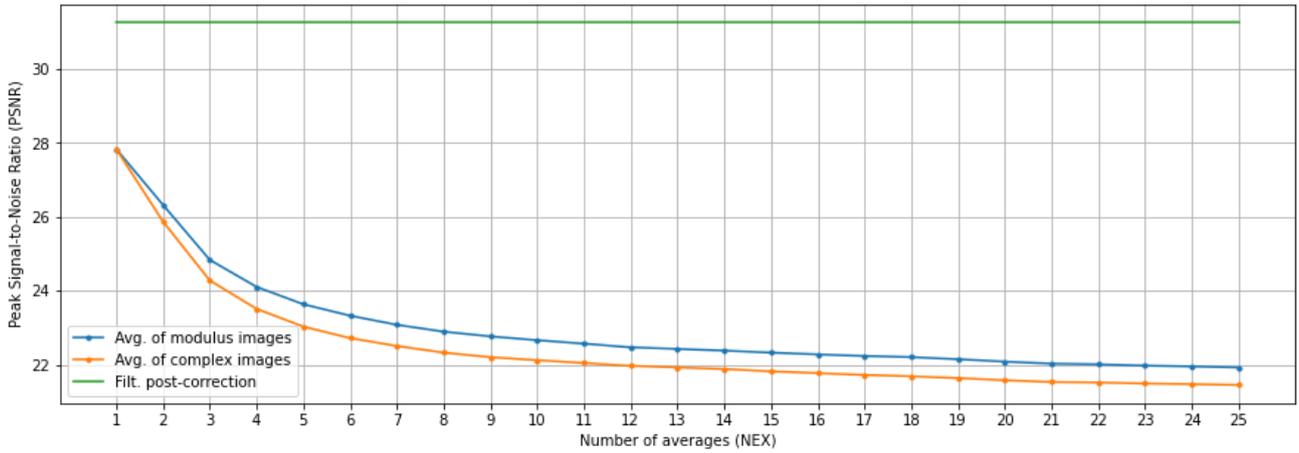

**Figure A24.** PSNR plot versus the NEX for the averaged images with SNR = 7. Pre- and post-correction denoised NEX = 1 image PSNR vales (for tissue) are shown as horizontal lines. Model with without PSF and motion.

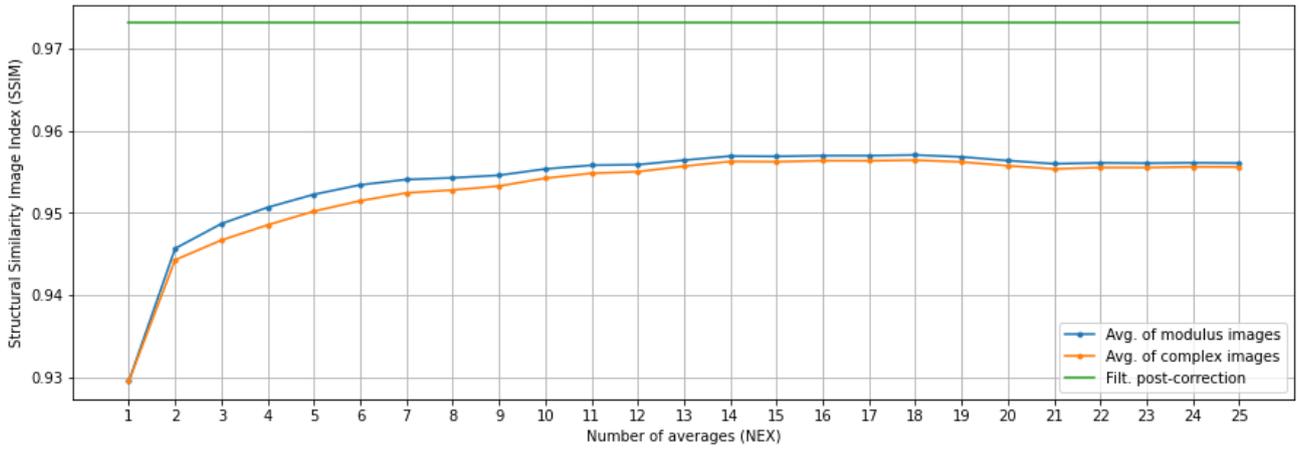

**Figure A25.** SSIM plot versus the NEX for the averaged images with SNR = 7. Pre- and post-correction denoised NEX = 1 image SSIM vales (for tissue) are shown as horizontal lines. Model with without PSF and motion.

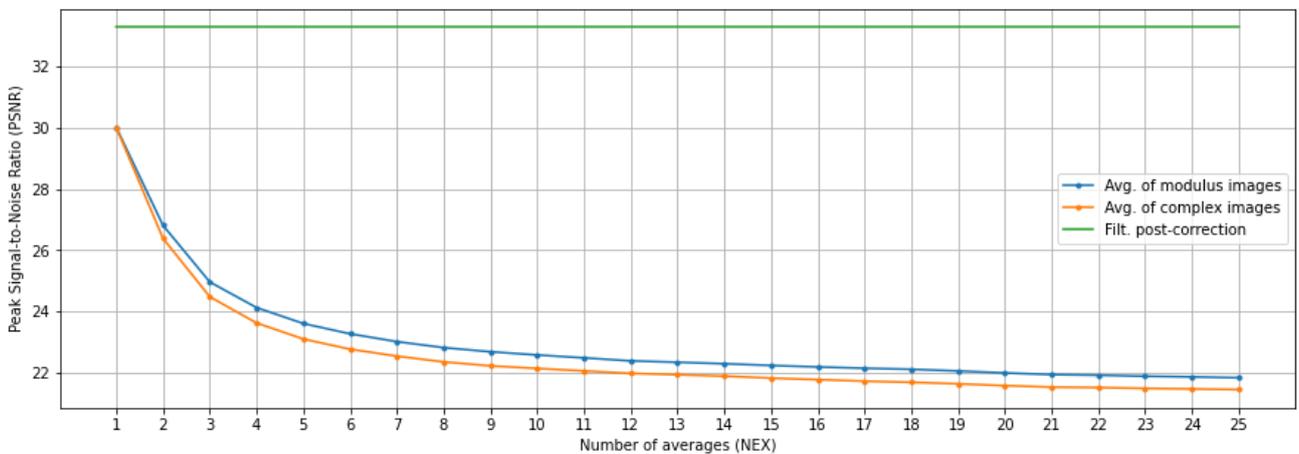

**Figure A26.** PSNR plot versus the NEX for the averaged images with SNR = 9. Pre- and post-correction denoised NEX = 1 image PSNR vales (for tissue) are shown as horizontal lines. Model with without PSF and motion.

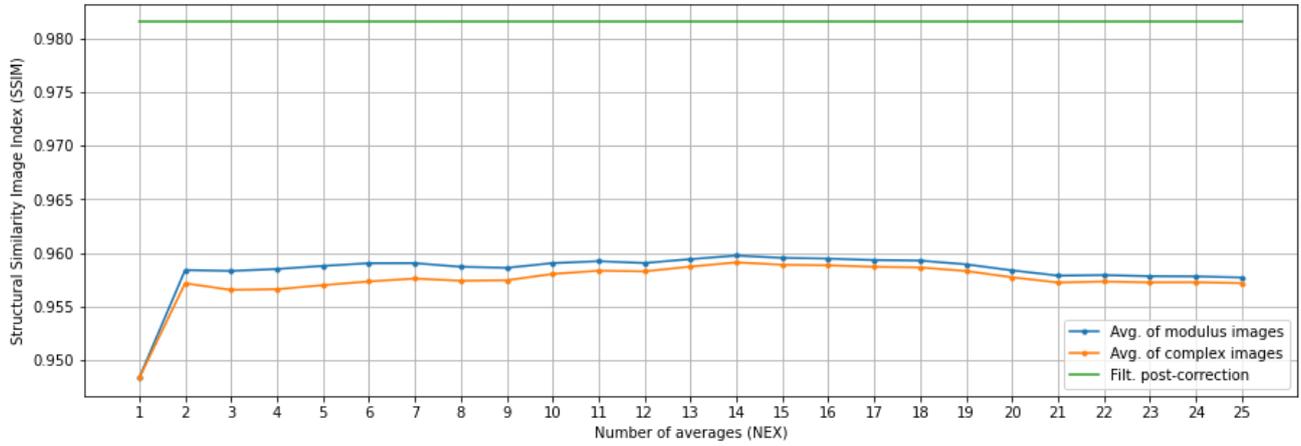

**Figure A27.** SSIM plot versus the NEX for the averaged images with SNR = 9. Pre- and post-correction denoised NEX = 1 image SSIM vales (for tissue) are shown as horizontal lines. Model with without PSF and motion.

Averaged images with varying SNR and NEX value – to Section 4.1.3.

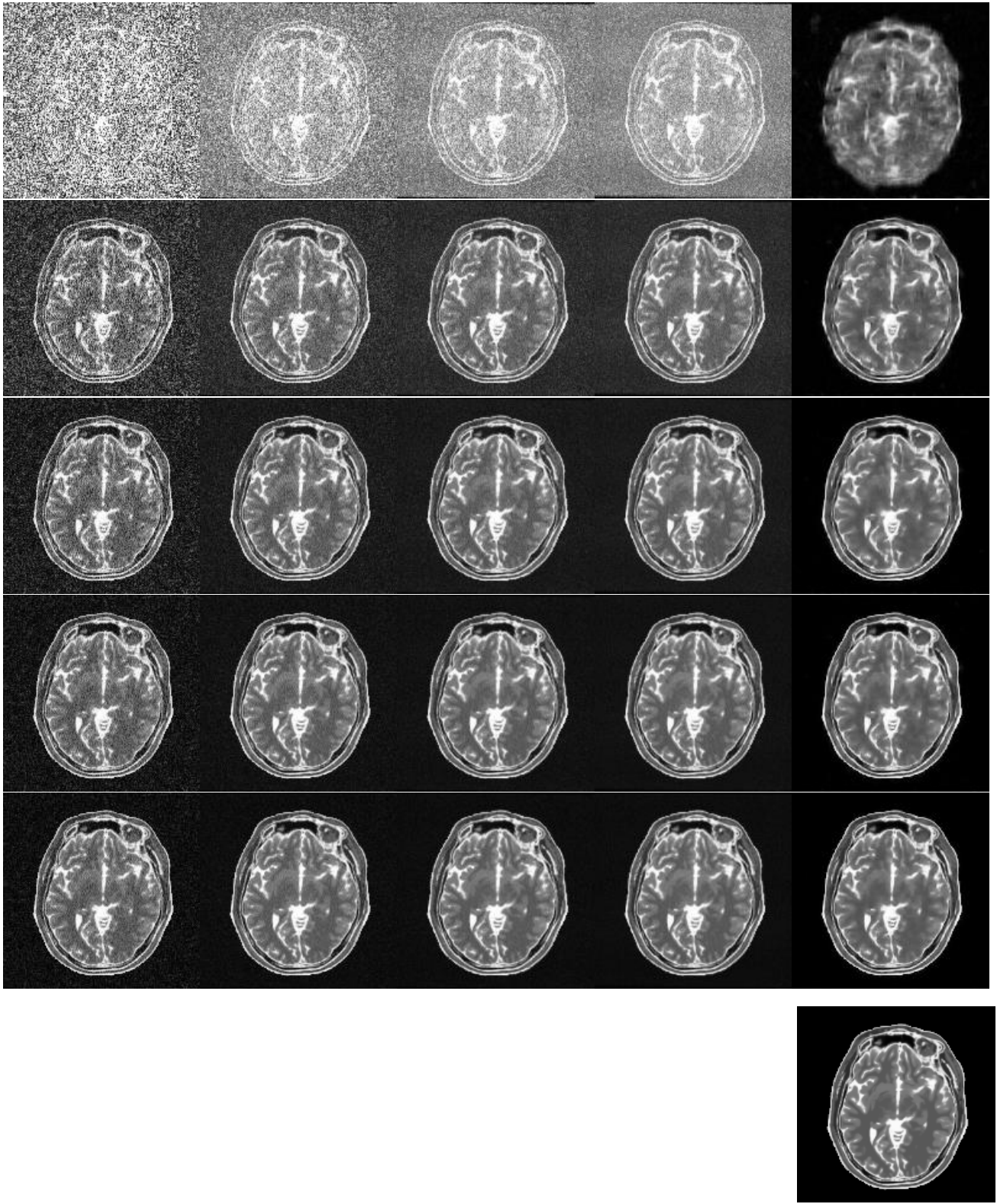

**Figure A28.** From left: Repeated images (ARI) averaged in the modulus domain: NEX=1, 4, 9, 16, then a CNN-denoised image. From top to bottom: SNR=1, 3, 5, 7, 9, last row: clean image. Model without PSF and with motion

Comparison of averaged images with varying NEX value to denoised images – to Section 4.2.

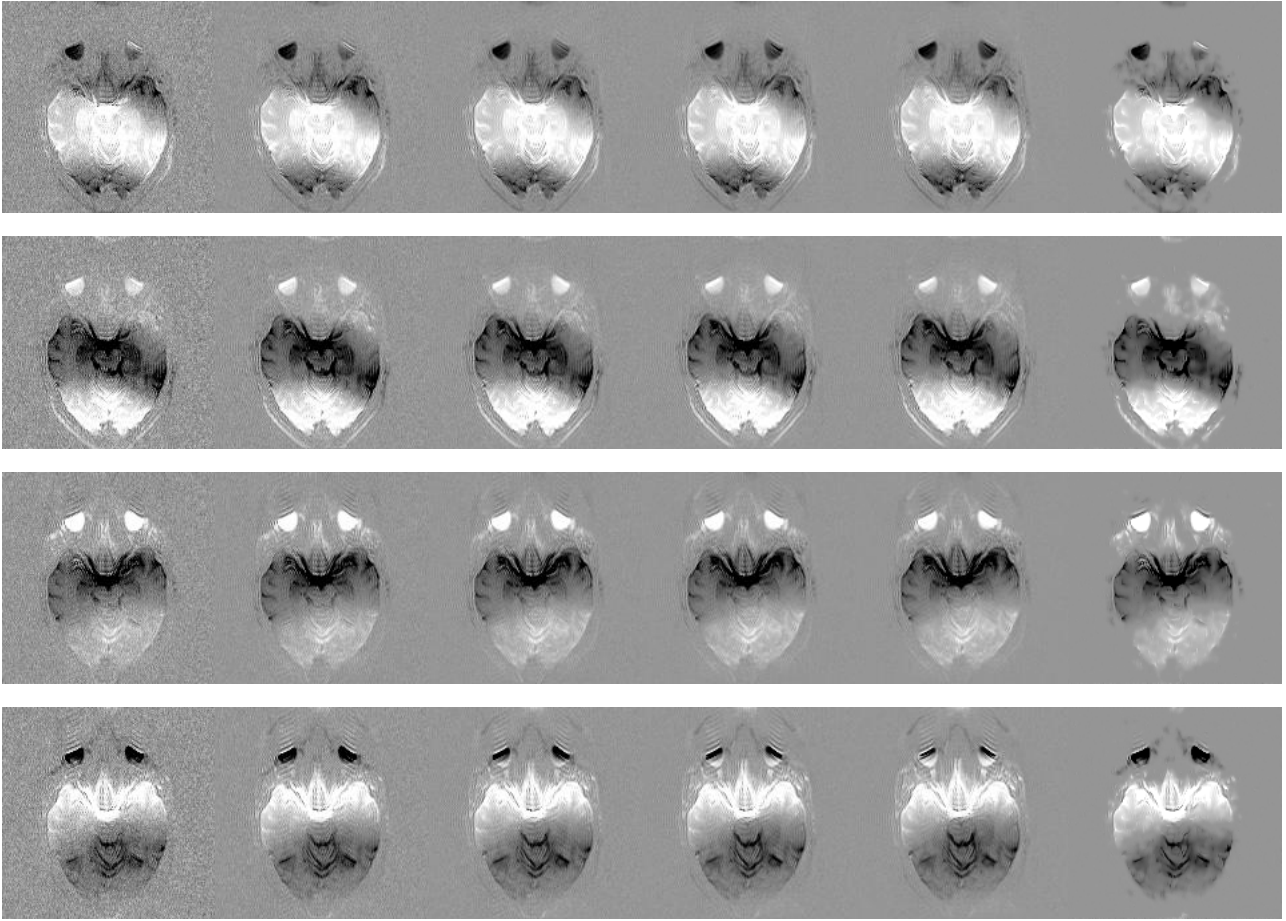

**Figure A29.** Denoising results of the CNN trained with SNR=7 simulated data, compared to averaged images (in the complex-image domain), for the case of a real component of a slice from b=0 DWI. From left to right: noisy image NEX=1, 4, 9, 16, 25, denoised image. From top to bottom: images from individual receiver coils used in the imaging experiment

Additional figures to Section 4.3.

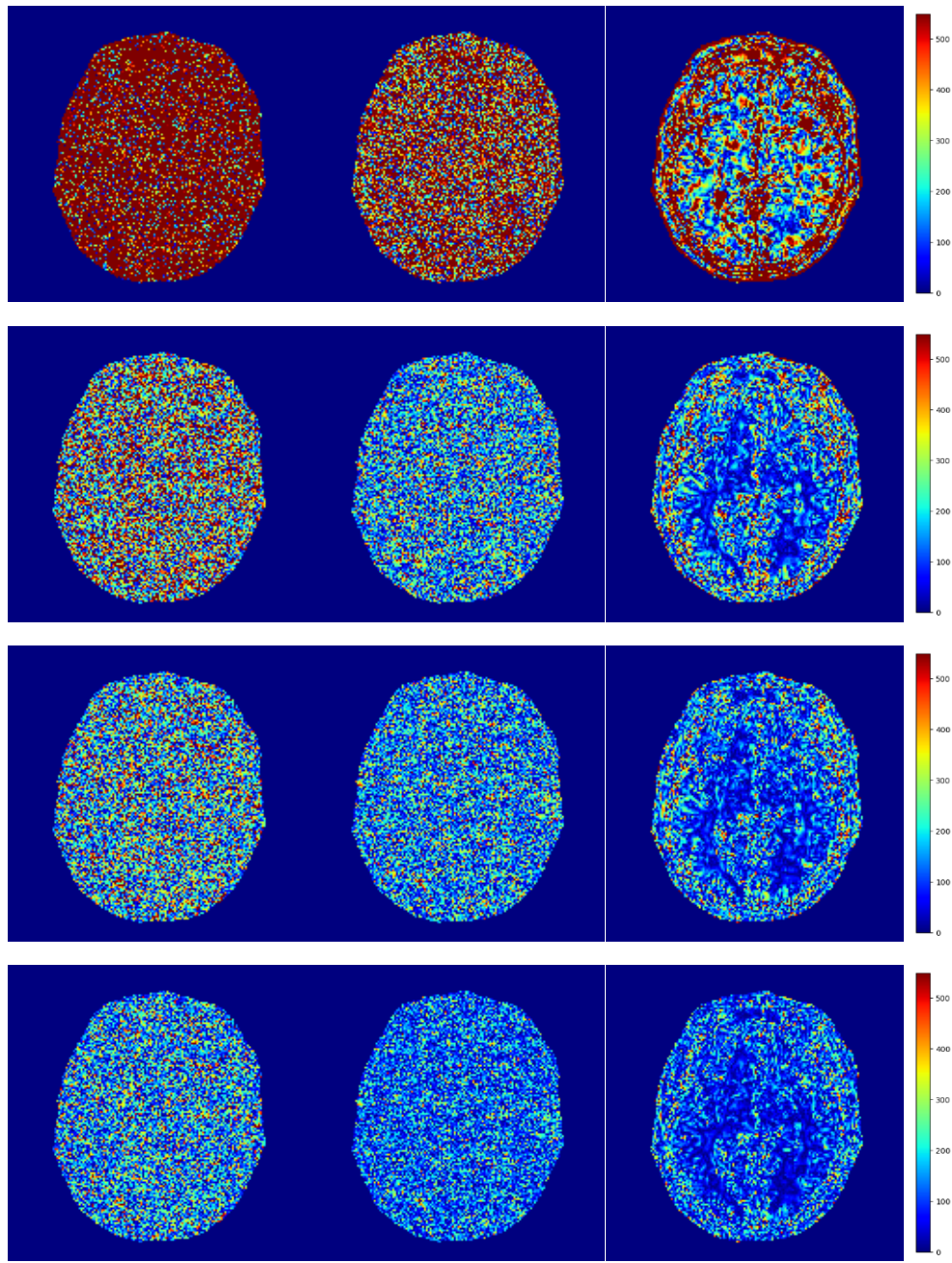

**Figure A30. From left to right: NEX=1 noisy image, NEX=3 averaged image, CNN-denoised image. From top to bottom: SNR = 1, 5, 7, 9. Model without the PSF and motion**

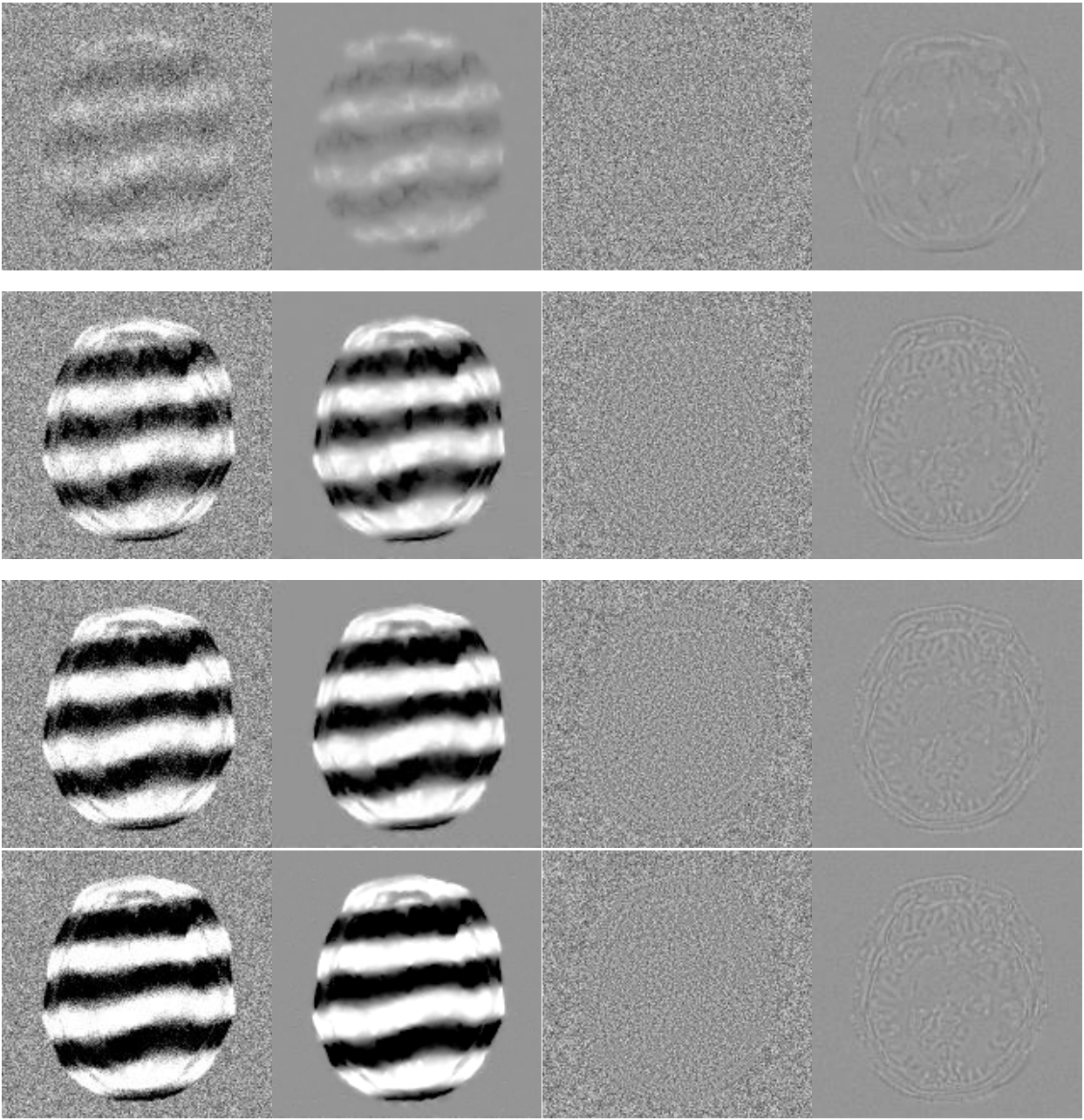

**Figure A31.** From left to right: noisy image (NEX=1), CNN-denoised image, estimated noise map, average of 100 estimated noise maps. Simulated image, real part, SNR values from top to bottom: 1, 5, 7, 9.

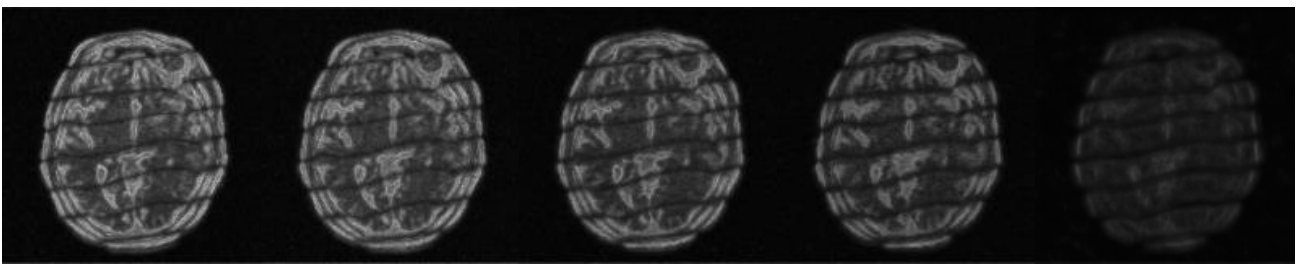

**Figure A32.** Standard deviation maps of the real components of the CNN-denoised image, from left: SNR = 9, 7, 5, 3, 1. Wavy lines in the middle image reflect phase changes in the noisy image, which cause intensity drops.

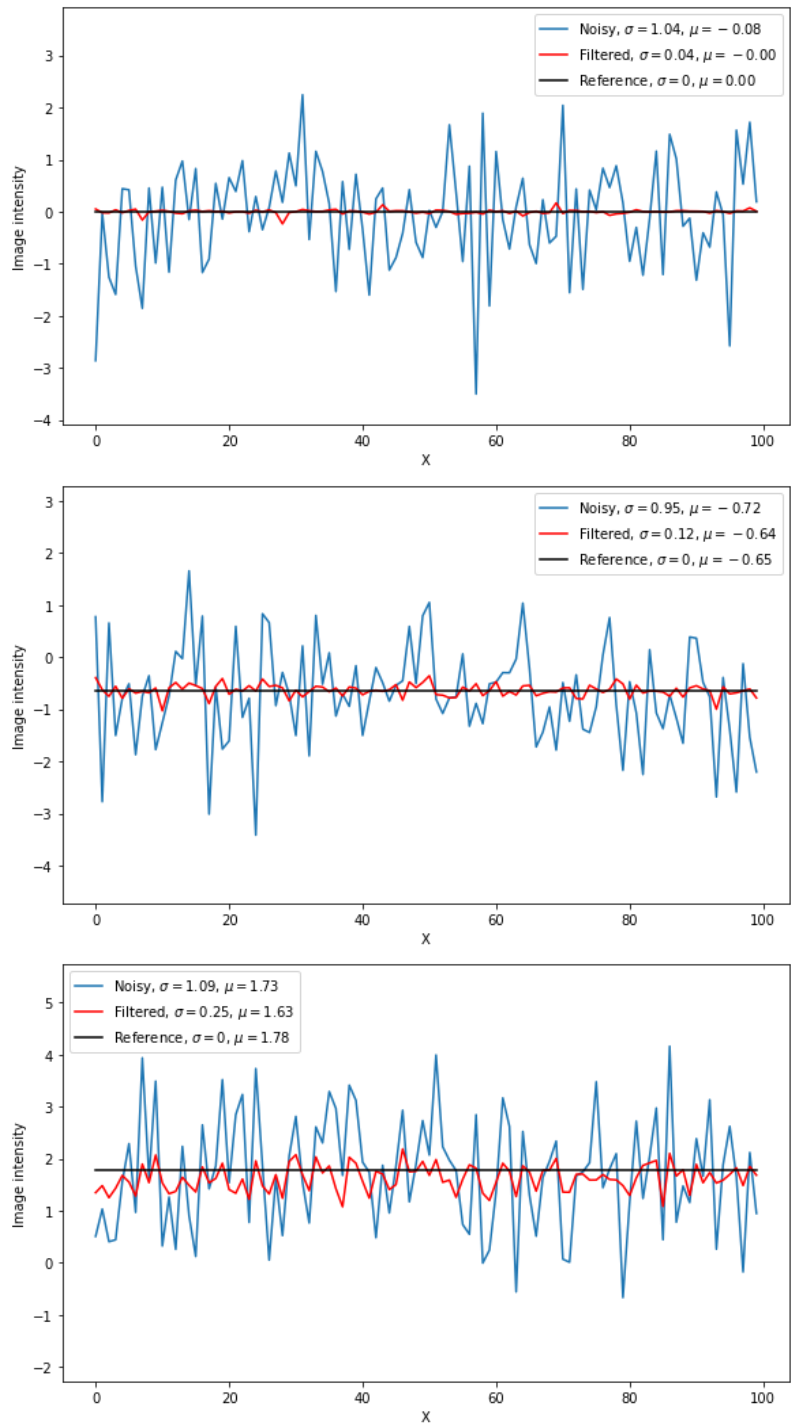

**Figure A33. Intensity plots for three chosen voxels: one located in the background, one in the mostly homogeneous area of the white matter and one in the high-intensity edge, across 100 noisy instances, SNR = 1. Noisy: blue line, denoised: red line.**

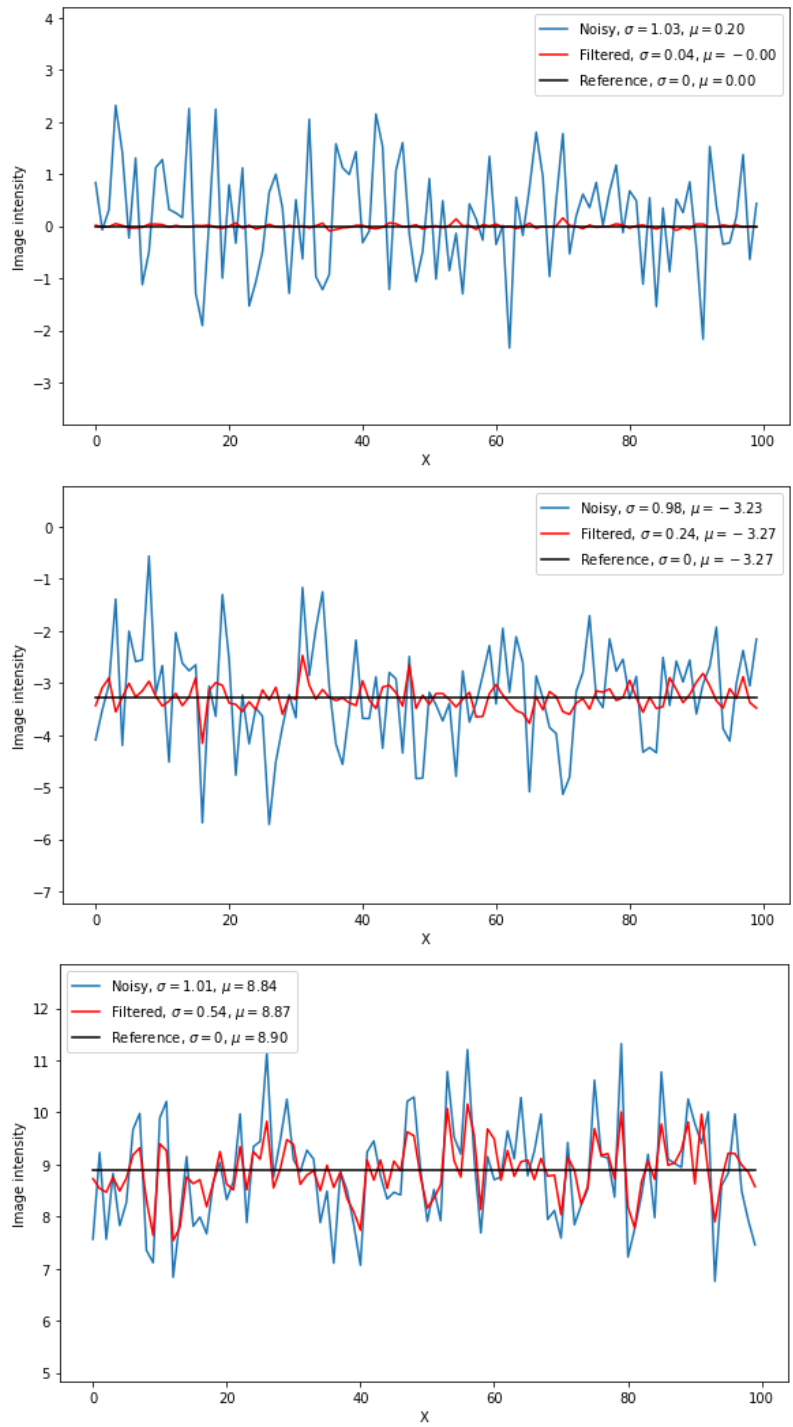

**Figure A34. Intensity plots for three chosen voxels: one located in the background, one in the mostly homogeneous area of the white matter and one in the high-intensity edge, across 100 noisy instances, SNR = 5. Noisy: blue line, denoised: red line.**

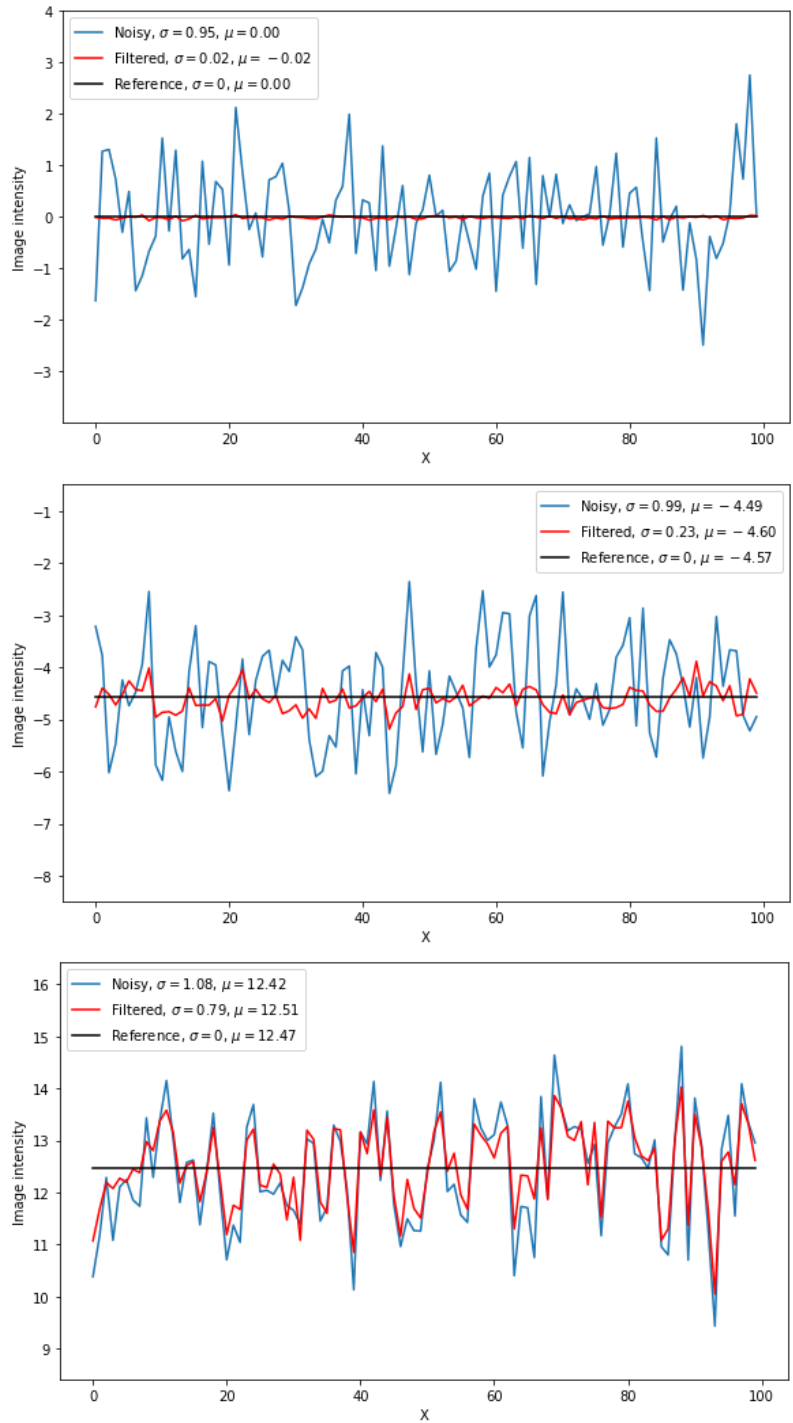

**Figure A35. Intensity plots for three chosen voxels: one located in the background, one in the mostly homogeneous area of the white matter and one in the high-intensity edge, across 100 noisy instances, SNR = 7. Noisy: blue line, denoised: red line.**

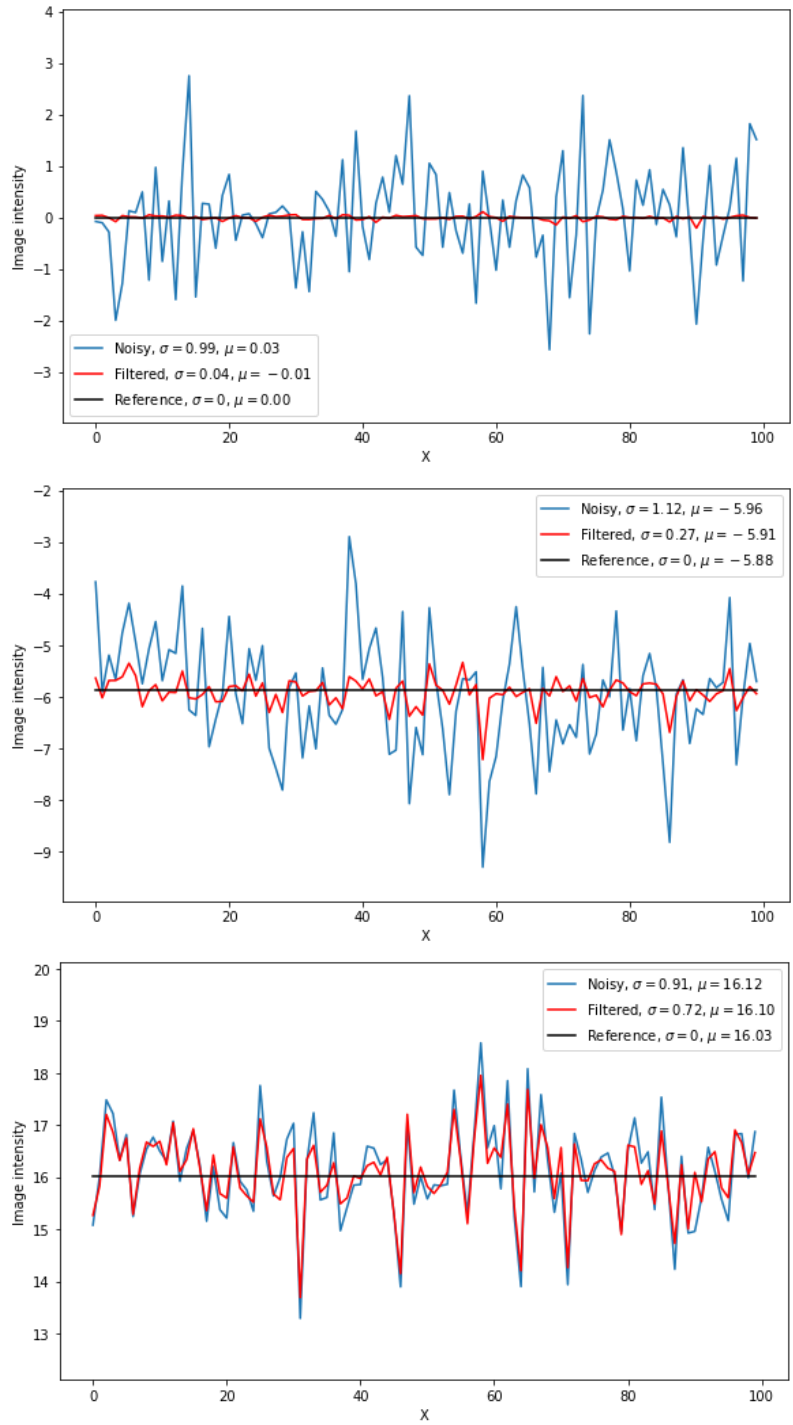

**Figure A36. Intensity plots for three chosen voxels: one located in the background, one in the mostly homogeneous area of the white matter and one in the high-intensity edge, across 100 noisy instances, SNR = 9. Noisy: blue line, denoised: red line.**